%% file: neurips_2026.tex
\documentclass{article}

\usepackage[main,final]{neurips_2026} 

\usepackage[utf8]{inputenc} 
\usepackage[T1]{fontenc}    
\usepackage{hyperref}       
\usepackage{url}            
\usepackage{booktabs}       
\usepackage{amsfonts}       
\usepackage{nicefrac}       
\usepackage{microtype}      
\usepackage{xcolor}         

\usepackage{amsmath}
\usepackage{amssymb}
\usepackage{mathtools}
\usepackage{amsthm}
\usepackage{thmtools, thm-restate}

\input{math_commands.tex}

\usepackage{graphicx}
\usepackage{subcaption}
\usepackage{wrapfig}
\usepackage{float}
\usepackage{multirow}
\usepackage{tabularray}
\usepackage{enumitem}
\usepackage{algorithm}
\usepackage{algpseudocode}

\title{Designing Cell-Type-Specific Regulatory DNA \newline with Guided Discrete Diffusion}

\author{%
  \textbf{Animesh Awasthi\textsuperscript{1,2,3} \quad Martin Stoll\textsuperscript{1,2,3} \quad Raphael Bednarsky\textsuperscript{1,2,3}}\\
  \textbf{Moritz Schaefer\textsuperscript{1,2,3} \quad Christoph Bock\textsuperscript{1,2,3}}\thanks{Correspondence to: Christoph Bock, \texttt{christoph.bock@meduniwien.ac.at}.}\\[0.5em]
  \textsuperscript{1}Medical University of Vienna, Institute of Artificial Intelligence,\\
  Center for Medical Data Science, Vienna, Austria\\[0.3em]
  \textsuperscript{2}CeMM Research Center for Molecular Medicine\\
  of the Austrian Academy of Sciences, Vienna, Austria\\[0.3em]
  \textsuperscript{3}Medical University of Vienna,\\
  Comprehensive Center for AI in Medicine (CAIM), Vienna, Austria
}
\begin{document}

\maketitle
\begin{abstract}
AI-driven design of DNA with cell-type-specific regulatory activity is broadly relevant for cell engineering and gene therapy. DNA elements such as promoters and enhancers have gene-regulatory functions that are defined by their DNA sequence and follow a characteristic regulatory grammar. Deep learning models can effectively predict DNA regulatory activity across cell types, but current generative methods often produce DNA sequences that deviate from the native regulatory grammar, increasing the risk of unexpected behavior and off-target effects in real-world applications. Here, we introduce DNA-CRAFT, a method for designing regulatory DNA with high predicted cell-type-specific activity while preserving naturalness. DNA-CRAFT formulates regulatory sequence design as a guided search over a generative model pre-trained on millions of naturally occurring regulatory elements from the human and mouse genomes. Our method enables targeted exploration of the DNA sequence space, efficient evaluation of candidate designs, and explicit suppression of unwanted activity. Across benchmarks spanning human cell lines and primary immune cells, DNA-CRAFT combines high predicted cell-type specificity with strong motif and sequence-composition adherence, achieving favorable results across the evaluated DNA design tasks.
\end{abstract}

\section{Introduction}
\label{sec:intro}

Precise control of gene expression is at the heart of programmable biology, and offers transformative potential for cell engineering and gene therapy \citep{dunbar_gene_2018, doi:10.1126/science.aad1067, yang_enhancer_2025}. A key task is designing regulatory DNA that drives high levels of gene expression in desired cell types while minimizing expression in undesired cell types. Naturally occurring regulatory DNA provides ample evidence that cell-type-specific gene expression is feasible, with combinations of transcription factor binding sites (TFBSs) following the genome's regulatory grammar \citep{doi:10.1126/science.adf7044, mitra_single-cell_2024}. However, the catalog of naturally occurring regulatory elements \citep{moore_expanded_2026} is often insufficient for applications in cell engineering and gene therapy, highlighting the need for methods to design synthetic regulatory DNA elements with desired properties. 

A key criterion for such methods is that the designed sequences preserve naturalness, i.e., remain consistent with the characteristic TFBS usage that defines the genome's regulatory grammar. While previous work has shown that even random DNA can exhibit regulatory activity \citep{de2020deciphering}, mammalian gene regulation depends on combinations of TFBSs within a supportive DNA sequence and chromatin context \citep{mitra_single-cell_2024}. Methods that do not consider the naturalness of the generated DNA risk exploiting biases in predictive models rather than encoding robust regulatory logic \citep{vaishnav2022evolution}, and may fail to generalize \textit{in vivo} where chromatin state and cellular context strongly modulate activity \citep{yin_iterative_2025}.

Current approaches for regulatory sequence design rely on sequence-to-activity models \citep{enformer, linder_predicting_2025} to predict activity across cell types. Optimization-based approaches, such as gradient-guided search \citep{schreiber_programmatic_2025, linder2021fast} and reinforcement learning fine-tuning \citep{chen_ctrl-dna_2025, wang_fine-tuning_2025}, can maximize differential activity but may push generated sequences far from the distribution of naturally occurring regulatory DNA \citep{vaishnav2022evolution}. Generative approaches such as classifier-free guidance (CFG)-guided discrete diffusion \citep{sarkar_designing_2025} can preserve naturalness but provide limited steering toward differential activity across cell types. Similar limitations apply to other related work, which we discuss in Appendix~\ref{app:related_work}. Hence, our central question is: \textit{How can we maximize cell-type-specific regulatory activity while remaining close to the distribution of naturally occurring regulatory DNA?}

Regulatory DNA design can be approached through inference-time search over a pre-trained discrete diffusion model \citep{sahoo_simple_2024, lou2024discretediffusionmodelingestimating}, using its learned sequence distribution as a prior. This connects inference-time alignment \citep{uehara2025inferencetimealignmentdiffusionmodels} with Monte Carlo tree search (MCTS) \citep{uct} to guide diffusion sampling \citep{tang_peptune_2024, yoon2026montecarlotreediffusion}. Adapting this approach to regulatory DNA design requires three considerations. First, conditioning expansion on the desired regulatory class can focus the search on relevant regions of sequence space. Second, evaluating candidates through full reverse diffusion rollouts is computationally expensive. Third, the reward must explicitly favor target-cell activity while penalizing off-target activity. These considerations motivate a search framework tailored to cell-type-specific regulatory DNA design.

In this work, we introduce \textbf{DNA-CRAFT} (\textbf{DNA} \textbf{C}is-\textbf{R}egulatory \textbf{A}rchitecture \& \textbf{F}unction \textbf{T}uner), a method for designing regulatory DNA with high predicted cell-type-specific activity while preserving naturalness. DNA-CRAFT combines a discrete diffusion model trained on more than 3.2~million naturally occurring regulatory elements from the human and mouse genomes \citep{moore_expanded_2026} with guided search over its sampling trajectory. The search incorporates three advances: regulatory-class-conditioned expansion via discrete CFG, fast lookahead pruning to efficiently evaluate candidate continuations, and a MinGap specificity objective with a smooth $\mathrm{LogSumExp}$ penalty on off-target activity. Across three human cell lines and eight primary immune cell types, DNA-CRAFT achieves a favorable trade-off between predicted specificity and the measured naturalness metrics. We also provide ablation studies assessing the contributions of its three search components.

\begin{figure}[h] 
    \centering 
    \includegraphics[width=1.0\linewidth]{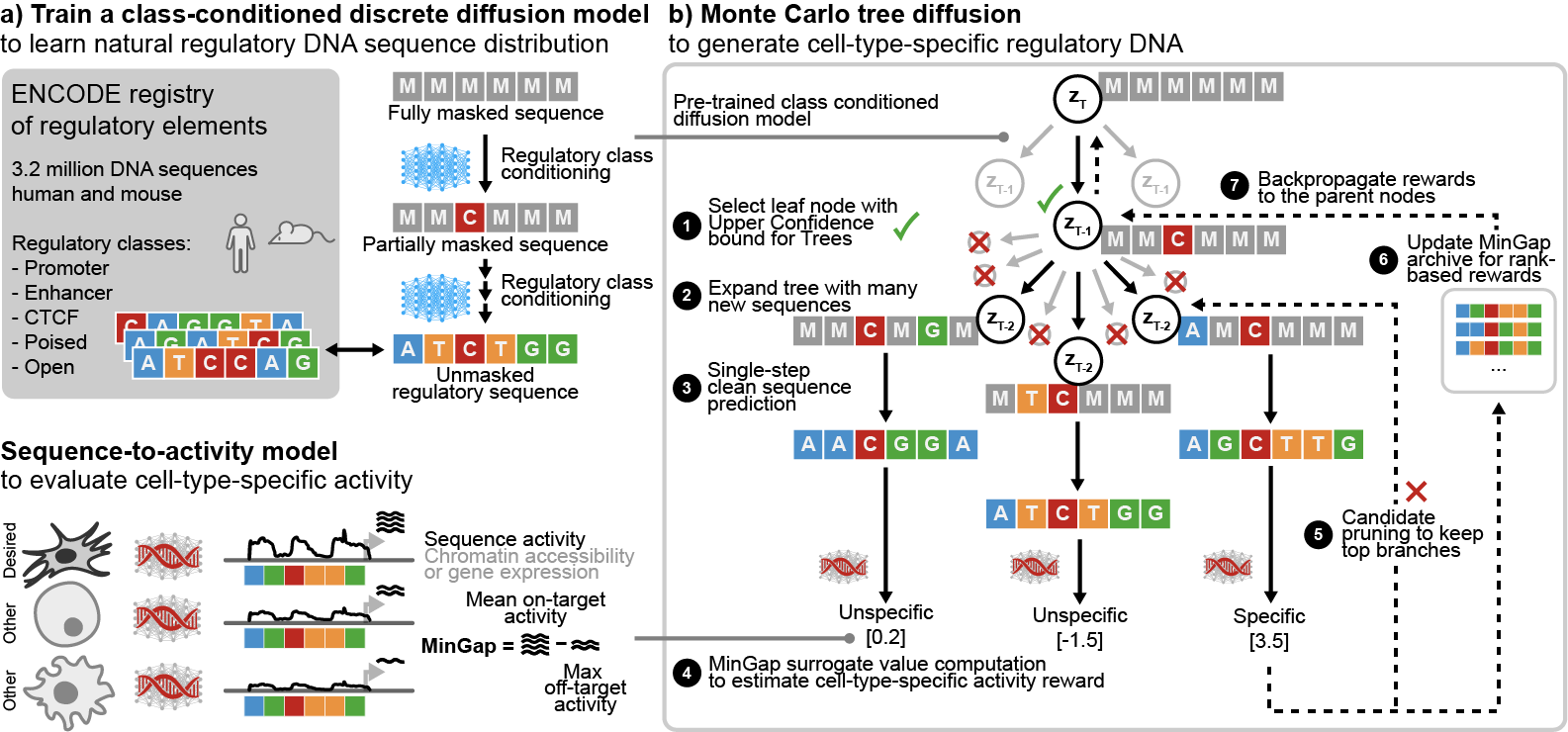} 
    \caption{\textbf{Overview of DNA-CRAFT.} \textbf{(a)} A class-conditioned discrete diffusion model is trained on the ENCODE registry of naturally occurring regulatory DNA. \textbf{(b)} At sampling time, Monte Carlo tree guidance steers the pre-trained class-conditioned diffusion model using cell-type-specific rewards.}
    \label{fig:main_figure} 
    \vspace{20pt}
\end{figure}

Our key contributions are as follows:
\begin{itemize}[nosep]
    \item We formulate cell-type-specific regulatory DNA design as an inference-time search problem over a frozen discrete diffusion prior, optimizing differential activity while using the learned distribution of naturally occurring regulatory DNA to guide generation.
    \item We extend Monte Carlo tree diffusion to regulatory DNA design through class-conditioned expansion, fast lookahead pruning, and a MinGap-based specificity objective.
    \item We benchmark DNA-CRAFT across cell-type-specific design tasks spanning three human cell lines and eight immune cell types, demonstrating a favorable trade-off between predicted specificity and the measured naturalness metrics.
\end{itemize}

Source code is available at \url{https://github.com/epigen/dna-craft}.

\section{Background and Problem Formulation}
\label{sec:background}

We model regulatory DNA with masked discrete diffusion \citep{sahoo_simple_2024}. Our implementation of the diffusion model is described in Appendix \ref{app:diffusion_training}.
Let $\mathcal{V} = \{A, C, G, T\}$ denote the token vocabulary and $\bm{m}$ denote the mask token (used only during the diffusion process). Each token is represented as a one-hot vector in $\mathbb{R}^{|\mathcal{V}|}$. Unmasked DNA sequences are denoted by $\bm{x} \in \mathcal{V}^L$, while partially masked sequences are denoted by $\bm{z}_t \in (\mathcal{V} \cup \{\bm{m}\})^L$ at continuous time $t \in [0, 1]$. We write $\bm{x} \sim \mathrm{Cat}(\bm{p})$ if $\bm{x}$ is drawn from a categorical distribution with parameter $\bm{p} \in \Delta^{|\mathcal{V}|}$. For sequences of length $L$, we write $\bm{x}^{1:L}$ and $\bm{z}_t^{1:L}$ (or $\bm{x}$ and $\bm{z}_t$ for brevity), with $\bm{x}^\ell$ and $\bm{z}_t^\ell$ denoting the $\ell$-th token.

\subsection{Problem Formulation: Cell-Type-Specific Regulatory DNA Sequence Design}
\label{problem_formulation}
The primary objective in regulatory sequence design is to generate a DNA sequence $\bm{x}^{1:L}$ that drives high gene expression (and/or chromatin accessibility) in a set of desired cell types while minimizing activity in a set of undesired cell types. To formalize this objective, we assume access to a pre-trained sequence-to-activity model that predicts regulatory activity across several cell types.

Let $\mathcal{C}$ denote the set of all evaluated cell types, and let $\bm{s}(\bm{x}^{1:L}) \in \mathbb{R}^{|\mathcal{C}|}$ be the activity vector predicted by the sequence-to-activity model for a sequence $\bm{x}^{1:L}$. We partition $\mathcal{C}$ into a set of target cell types $\mathcal{C}^+$ and a set of undesired background cell types $\mathcal{C}^-$ and formulate cell-type specificity through two aggregation operators acting on the predicted activities over target and off-target cell types. Let $A^+$ and $A^-$ denote aggregation operators over $\mathcal C^+$ and $\mathcal C^-$, respectively. The reward is defined as
\[
r(\bm{x}^{1:L}) =
A^+\!\left(\{s_c(\bm{x}^{1:L})\}_{c\in\mathcal C^+}\right)
-
A^-\!\left(\{s_c(\bm{x}^{1:L})\}_{c\in\mathcal C^-}\right).
\]
A common choice in prior work is to use mean aggregation over both target and off-target sets, effectively optimizing average activity differences \citep{reddy2024designing}. However, such formulations do not explicitly control for individual off-target cell types, and may allow sequences that exhibit high activity in specific unintended contexts. In many biological applications, even moderate activity in a single off-target cell type can be undesirable, motivating a more conservative notion of specificity. To address this, we adopt a worst-case formulation over off-target activity, leading to the \textbf{MinGap} score \citep{gosai2024machine}. Concretely, we set $A^+=\mathrm{Mean}$ and $A^-=\mathrm{Max}$, yielding
\begin{equation}
\label{eq:mingap}
    r(\bm{x}^{1:L}) = \underbrace{\frac{1}{|\mathcal{C}^+|} \sum_{c \in \mathcal{C}^+} s_c(\bm{x}^{1:L})}_{\text{Mean Activity (Target)}} 
- \underbrace{\max_{c \in \mathcal{C}^-} s_c(\bm{x}^{1:L})}_{\text{Max Activity (Off-target)}}.
\end{equation}
This objective encourages high target activity relative to the largest predicted off-target response. For search, we replace the maximum with a smooth $\mathrm{LogSumExp}$ aggregation and define
\begin{equation}
\label{eq:smooth_mingap}
g(\bm{x}^{1:L}) = \frac{1}{|\mathcal{C}^{+}|}\sum_{c\in\mathcal{C}^{+}}s_c(\bm{x}^{1:L}) - \log\!\left(\sum_{c\in\mathcal{C}^{-}}\exp(s_c(\bm{x}^{1:L}))\right).
\end{equation}
The LogSumExp uses unit temperature without normalization by the number of off-target cell types.

\subsection{Inference-Time Alignment and the Specificity-Naturalness Trade-off}
We aim to maximize the specificity reward $r(\bm{x}^{1:L})$ while remaining consistent with the pre-trained diffusion model, leading to the following inference-time alignment objective~\citep{uehara2025inferencetimealignmentdiffusionmodels}:

\begin{equation}
\tilde{p}(\bm{x}^{1:L}) = \arg\max_{p} \; \mathbb{E}_{\bm{x}^{1:L} \sim p} \bigl[ r(\bm{x}^{1:L}) \bigr] - \lambda D_{\mathrm{KL}}(p \,\|\, p_\theta)
\;\propto\;
p_\theta(\bm{x}^{1:L}) \cdot \exp\!\left(\frac{r(\bm{x}^{1:L})}{\lambda}\right),
\end{equation}
where $\lambda > 0$ controls the trade-off between reward maximization and adherence to the generative prior $p_\theta$. Small $\lambda$ concentrates the distribution on high-reward sequences, while large $\lambda$ recovers the prior. Navigating this trade-off is the central challenge in regulatory DNA sequence design.

\textbf{Soft value function.} The target distribution $\tilde{p}$ is defined over fully denoised sequences $\bm{x}^{1:L}$, while diffusion proceeds through intermediate states $\bm{z}_t$. To guide this process, we require a measure of the desirability of each intermediate state. This is captured by the \emph{soft value function}, which estimates the expected terminal reward from a given state:
\begin{equation}
\label{eq:VF}
v_t(\bm{z}_t^{1:L}) \coloneqq \lambda \log \mathbb{E}_{\bm{x}^{1:L} \sim p_\theta(\cdot \mid \bm{z}_t^{1:L})}
\left[\exp\left(\frac{r(\bm{x}^{1:L})}{\lambda}\right)\right],
\end{equation}
which aggregates the rewards of all possible completions of $\bm{z}_t^{1:L}$, emphasizing high-reward outcomes according to $\lambda$.

\textbf{Soft optimal policy.} The value function induces the soft optimal policy \citep{uehara2025inferencetimealignmentdiffusionmodels}, which defines the ideal reverse transition for $ s < t $:
\begin{equation}
\label{eq:optimal_policy}
p^\star_s(\bm{z}_s^{1:L} \mid \bm{z}_t^{1:L}) 
\;=\; 
\frac{
  p_\theta(\bm{z}_s^{1:L} \mid \bm{z}_t^{1:L}) 
  \;\exp\!\bigl(v_s(\bm{z}_s^{1:L}) / \lambda\bigr)
}{
  \exp\!\bigl(v_t(\bm{z}_t^{1:L}) / \lambda\bigr)
}.
\end{equation}
This policy retains the structure of the pre-trained denoising distribution $p_\theta$ while re-weighting each candidate continuation $\bm{z}_s^{1:L}$ by its value $v_s$. Sequentially sampling from $p^\star_s$ at every denoising step produces sequences distributed according to the target $\tilde{p}$. However, sampling from $p^\star_s$ is intractable for two reasons: (i)~evaluating $v_t$ requires marginalizing over all possible completions of $\bm{z}_t^{1:L}$, an exponentially large space for DNA sequences; and (ii)~computing the normalizing constant in the denominator of the soft optimal policy is equally prohibitive.

\textbf{Approximation strategy.} We approximate the soft optimal policy using Monte Carlo Tree Diffusion (MCTD)~\citep{tang_peptune_2024, yoon2026montecarlotreediffusion, yoon2025fastmontecarlotree}, which builds a search tree over intermediate states $\bm{z}_t^{1:L}$ and uses surrogate value estimates to guide expansion toward high-reward regions while remaining consistent with the generative prior. We note that this results in a heuristic search procedure rather than an exact sampler from the KL-regularized target distribution.

\subsection{Monte Carlo Tree Diffusion For Regulatory DNA Sequences}
\label{subsec:mctd}

We formulate regulatory DNA design as a search over partially denoised diffusion states. Nodes correspond to sequences $\bm{z}_t$, and edges represent reverse transitions to $\bm{z}_s$ with $s<t$. The maximum depth of the resulting tree matches the number of denoising steps. The number of candidate children sampled per expansion at each node is $M$. MCTD for discrete diffusion undergoes four phases:
\begin{enumerate}[leftmargin=*]
    \item \textbf{Selection:} The algorithm traverses from the root to a leaf node by selecting branches that maximize an Upper Confidence Bound for Trees (UCT) \citep{uct}, balancing the exploitation of high-reward trajectories with the exploration of the generative prior $p_\theta$.
    \item \textbf{Expansion:} Upon reaching a leaf node at time $t$, the tree branches out by sampling $M$ child nodes ($\bm{z}_{s,1}^{1:L}, \dots, \bm{z}_{s,M}^{1:L}$) from the reverse transition $p_\theta(\bm{z}_s^{1:L}\mid\bm{z}_t^{1:L})$, without regulatory-class conditioning in this standard formulation.
    \item \textbf{Rollout (Simulation):} Each child is then iteratively denoised all the way to an unmasked sequence ($t=0$) using the pre-trained diffusion model, and is scored by the sequence-to-activity model to obtain a Pareto reward (a multi-objective score based on dominance across cell types).
    \item \textbf{Backpropagation:} The calculated rewards are passed back up the tree, updating the cumulative rewards and visit counts of all parent nodes to inform future selection steps.
\end{enumerate}

\section{Methods}
\label{sec:methods}
DNA-CRAFT builds on the tree-search procedure described above to guide a pre-trained discrete diffusion model toward cell-type-specific regulatory DNA. We first describe the class-conditioned generative prior, then explain how regulatory-class conditioning, fast lookahead pruning, and the MinGap specificity reward guide candidate generation, selection, and tree updates (Figure~\ref{fig:main_figure}). The complete algorithm is provided in Appendix~\ref{app:complete_algorithm}.

\subsection{Class-Conditioned Discrete Diffusion for Regulatory DNA}
To provide a biologically grounded generative prior, we train a class-conditioned masked diffusion language model on the ENCODE registry of candidate regulatory elements, comprising over 3.2 million naturally occurring DNA sequences across the human and mouse genomes \citep{moore_expanded_2026}. We use a bidirectional Mamba \citep{gu2024mambalineartimesequencemodeling} state space model backbone with reverse-complement equivariance that preserves the double-stranded geometry of DNA \citep{schiff_caduceus_2024}. 

We train a class-conditioned denoising network, where $\bm{y}$ represents one of five consolidated regulatory element classes from the ENCODE registry. Let $\bm{x}_\theta(\bm{z}_t,t,\bm{y})$ denote its clean-nucleotide probabilities at all positions, and let $\bm{x}_\theta(\bm{z}_t,t)$ denote predictions from the same network with the null conditioning label. During inference, we apply discrete classifier-free guidance (CFG)~\citep{schiff_simple_2025} to obtain
\begin{equation}
\label{eq:cfg}
\bm{x}_\theta^\gamma(\bm{z}_t,t,\bm{y}) = \operatorname{softmax}\!\left(\gamma\log\bm{x}_\theta(\bm{z}_t,t,\bm{y}) + (1-\gamma)\log\bm{x}_\theta(\bm{z}_t,t)\right).
\end{equation}
For position $\ell\in\{1,\ldots,L\}$ and nucleotide $a\in\mathcal{V}$, $[\bm{x}_\theta^\gamma(\bm{z}_t,t,\bm{y})]_{\ell,a}$ denotes the predicted probability of nucleotide $a$ at position $\ell$. The logarithm acts element-wise, and softmax normalizes over nucleotides at each masked position. These guided clean-token probabilities are used to construct the reverse transition $p_\theta^\gamma(\bm{z}_s^{1:L}\mid\bm{z}_t^{1:L},\bm{y})$, as detailed in Appendix~\ref{app:cfg_renorm}. Unmasked positions remain unchanged and the denoising network remains frozen during search.

\subsection{Class-Conditioned Monte Carlo Tree Guidance}

Given the generative prior, we guide sampling using class-conditioned expansion, fast lookahead pruning, and an archive-relative specificity reward.

\textbf{1. Class-Conditioned Tree Expansion.} The genome contains distinct classes of regulatory elements with characteristic sequence composition and motif usage~\citep{friedman_enhancerpromoter_2024}. To focus search on the desired regulatory class, we sample candidate continuations from the CFG-guided reverse transition. 
When expanding a leaf node $\bm{z}_t^{1:L}$, we draw a pool of $B$ candidates and retain up to $M$ children after lookahead evaluation. Sampling uses the Gumbel-max construction and unmasked positions remain unchanged. Let $\bm{G}_b$ contain independent $\mathrm{Gumbel}(0,1)$ draws for every position and token of candidate $b$. At each masked position $\ell$, we sample
\begin{equation}
\label{eq:child_sampling}
\bm{z}_{s,b}^{\ell} = \operatorname*{arg\,max}_{a\in\mathcal{V}\cup\{\bm{m}\}}\left\{\log p_\theta^\gamma(\bm{z}_s^\ell=a\mid\bm{z}_t,\bm{y}) + [\bm{G}_b]_{\ell,a}\right\}.
\end{equation}

\textbf{2. Fast Lookahead Pruning.}

\textbf{Candidate pruning.} Inspired by value-based selection in inference-time alignment~\citep{uehara2025inferencetimealignmentdiffusionmodels, li_derivative-free_2024}, we evaluate the $B$ proposed continuations and retain the $M$ highest-scoring candidates, or all candidates when $B<M$. This concentrates further search on promising continuations within the sampled pool.

\textbf{Efficient value approximation.} Full reverse-diffusion rollouts are expensive when evaluating a large candidate pool. We instead construct a one-step tokenwise modal completion using the denoiser's clean-token predictions:
\begin{equation}
\label{eq:x0_map}
\hat{\bm{x}}_{0,b}^{\ell} =
\begin{cases}
\bm{z}_{s,b}^{\ell}, & \bm{z}_{s,b}^{\ell}\neq\bm{m},\\
\displaystyle\operatorname*{arg\,max}_{a\in\mathcal{V}}[\bm{x}_\theta^\gamma(\bm{z}_{s,b},s,\bm{y})]_{\ell,a}, & \bm{z}_{s,b}^{\ell}=\bm{m}.
\end{cases}
\end{equation}
We refer to this construction as the one-step MAP estimator, where MAP denotes the mode of the factorized clean-token prediction rather than the most likely completion under a full diffusion rollout. The surrogate value is $\hat{v}_s(\bm{z}_{s,b}^{1:L})=g(\hat{\bm{x}}_{0,b}^{1:L})$. Evaluation requires a one-step denoiser prediction followed by a sequence-to-activity model evaluation, avoiding a full reverse-diffusion rollout for each candidate. Appendix~\ref{app:map_validation} reports weak association with rollout scores at early steps and stronger association later in denoising.

\textbf{3. MinGap Specificity Reward.} We use the smooth specificity score $g$ from Equation~\ref{eq:smooth_mingap} to rank candidate completions and retain the highest-scoring children. For tree updates, we compare each retained completion with a bounded archive $\mathcal{G}^*$ of previously admitted record-scoring completions. Candidates are processed in decreasing score order, where the first candidate is inserted directly and subsequent candidates are admitted only when their score matches or exceeds every score currently in the archive. Previously admitted entries remain until the capacity $N_{\text{max}}$ is exceeded, at which point older strictly lower-scoring entries are removed first, with minimum-score removal used otherwise.

Before updating the archive for candidate $m$, we compute its relative reward:
\begin{equation}
\label{eq:rank_reward}
\rho_m =
\begin{cases}
1, & \mathcal{G}^*=\emptyset,\\
\displaystyle\frac{1}{|\mathcal{G}^*|}\sum_{\bm{x}^*\in\mathcal{G}^*}\mathbb{I}\!\left[g(\hat{\bm{x}}_{0,m}^{1:L})\geq g(\bm{x}^*)\right], & \text{otherwise}.
\end{cases}
\end{equation}
The reward lies between zero and one and measures the fraction of archived scores matched or exceeded by the candidate. Since the archive changes during search, these rewards are non-stationary. Candidate pruning uses the smooth score $g$, while tree selection uses accumulated archive-relative rewards. Complete initialization and backpropagation rules are provided in Appendix~\ref{app:complete_algorithm}.

\section{Experiments}
\label{sec:experiments}
\subsection{Human Cell-Line-Specific Enhancer Design}
\label{sec:human_cell_line}

\textbf{Experimental Setup.} We utilized a dataset of approximately 700,000 sequences with massively parallel reporter assay (MPRA) activity measurements across three cell lines \citep{gosai2024machine}. We employed a split-model validation strategy following \citet{wang_fine-tuning_2025}. To prevent data leakage, we partitioned the MPRA dataset into two halves based on sequence homology scores calculated using MMseqs2 \citep{steinegger_mmseqs2_2017}. We fine-tuned the Enformer model \citep{enformer} on the first half to create a Design-model and on the second half of the data to create the Evaluation-model. Results were further confirmed using independent evaluations (Table \ref{tab:human_enhancer_extended}).

\textbf{Benchmarking methods.} We benchmarked DNA-CRAFT (conditioned on the ``Enhancer'' class, $\gamma=1.0$, lookahead pool $B=800$) against state-of-the-art sequence design methods that use inference-time alignment approaches, including sequential Monte Carlo (SMC) \citep{phillips_particle_2024} and twisted diffusion sampling (TDS) \citep{wu_practical_2024} adapted for the MinGap objective, as well as CFG-guided DNA discrete diffusion (D3) \citep{sarkar_designing_2025}. We also compared against reinforcement learning fine-tuning techniques including DRAKES \citep{wang_fine-tuning_2025} for discrete diffusion and Ctrl-DNA \citep{chen_ctrl-dna_2025} for autoregressive models and the gradient-based optimization method Ledidi \citep{schreiber_programmatic_2025}. All methods used the same Design-model. Further implementation details are provided in Appendix \ref{app:benchmark_details}.

\textbf{Evaluation Metrics.} We use the MinGap score to assess cell-type specificity, two complementary metrics to assess naturalness and a metric to assess the diversity of the designed sequences. Details of the four metrics are: (\romannumeral1) \textbf{MinGap Score.} We compute the MinGap score (Equation  \ref{eq:mingap}) to measure differential activity in the desired cell type using the Evaluation-model. (\romannumeral2) \textbf{Motif Correlation.} For each desired cell type, we select reference sequences at or above the 99.9th percentile of MinGap scores computed from measured MPRA activity, corresponding to approximately the top 0.1\% of sequences. We scan these top sequences with FIMO \citep{grant_fimo_2011} and the JASPAR 2024 core vertebrate database \citep{rauluseviciute_jaspar_2024} to compute TFBS frequency distribution. This serves as the reference for evaluating the biological fidelity of our designed sequences. We compute the Spearman correlation between TFBS frequencies in the generated sequences and the reference. Higher correlation indicates closer agreement with the reference TFBS frequency distribution. (\romannumeral3) \textbf{3-mer Correlation.} We compute the Pearson correlation of 3-mer counts between generated and top reference sequences, capturing sequence composition. (\romannumeral4) \textbf{Diversity.} We calculate the mean per-position Shannon entropy across the generated batch using base-2 logarithms. This metric measures marginal nucleotide diversity, with a maximum of 2 bits per position.
\begin{table}[h]
\centering
\vspace{-10pt}
\caption{Comparison of methods for designing cell-type-specific enhancers across three human cell lines. Shown are the MinGap specificity scores (evaluated under a held-out Evaluation oracle), Motif Spearman correlations, 3-mer Pearson correlations, and Shannon diversity. Values are reported as mean $\pm$ std over 3 independent runs and the best result is marked in \textbf{bold}.}
\label{tab:enhancer_results}
\resizebox{\textwidth}{!}{%
\begin{tabular}{llccccccc}
\toprule
\textbf{Cell Line} & \textbf{Metric} & \textbf{SMC} & \textbf{TDS} & \textbf{D3} & \textbf{DRAKES} & \textbf{Ctrl-DNA} & \textbf{Ledidi} & \textbf{DNA-CRAFT (Ours)} \\
\midrule
\multirow{4}{*}{\textbf{HepG2}} 
 & MinGap Eval $\uparrow$ & 0.989 \scriptsize{$\pm 1.717$} & 0.880 \scriptsize{$\pm 1.053$} & -0.075 \scriptsize{$\pm 0.021$} & -1.314 \scriptsize{$\pm 0.024$} & \textbf{7.464 \scriptsize{$\pm 0.270$}} & 6.712 \scriptsize{$\pm 0.270$} & 7.271 \scriptsize{$\pm 0.142$} \\
 & Motif Corr. $\uparrow$ & 0.302 \scriptsize{$\pm 0.107$} & 0.456 \scriptsize{$\pm 0.152$} & 0.793 \scriptsize{$\pm 0.009$} & -0.126 \scriptsize{$\pm 0.014$} & 0.545 \scriptsize{$\pm 0.063$} & 0.522 \scriptsize{$\pm 0.019$} & \textbf{0.810 \scriptsize{$\pm 0.028$}} \\
 & 3-mer Corr. $\uparrow$ & 0.541 \scriptsize{$\pm 0.126$} & 0.735 \scriptsize{$\pm 0.083$} & \textbf{0.977 \scriptsize{$\pm 0.006$}} & -0.361 \scriptsize{$\pm 0.012$} & 0.568 \scriptsize{$\pm 0.098$} & 0.711 \scriptsize{$\pm 0.010$} & 0.930 \scriptsize{$\pm 0.003$} \\
 & Shannon Div. $\uparrow$ & 0.845 \scriptsize{$\pm 0.478$} & 1.301 \scriptsize{$\pm 0.105$} & 1.978 \scriptsize{$\pm 0.003$} & 1.864 \scriptsize{$\pm 0.002$} & 1.914 \scriptsize{$\pm 0.033$} & 1.980 \scriptsize{$\pm 0.002$} & \textbf{1.981 \scriptsize{$\pm 0.001$}} \\
\midrule
\multirow{4}{*}{\textbf{K562}} 
 & MinGap Eval $\uparrow$ & 1.812 \scriptsize{$\pm 0.768$} & 1.735 \scriptsize{$\pm 0.660$} & 0.003 \scriptsize{$\pm 0.070$} & -0.033 \scriptsize{$\pm 0.032$} & \textbf{8.490 \scriptsize{$\pm 0.153$}} & 8.004 \scriptsize{$\pm 0.078$} & 7.777 \scriptsize{$\pm 0.054$} \\
 & Motif Corr. $\uparrow$ & 0.269 \scriptsize{$\pm 0.130$} & 0.575 \scriptsize{$\pm 0.121$} & 0.794 \scriptsize{$\pm 0.057$} & 0.060 \scriptsize{$\pm 0.008$} & 0.528 \scriptsize{$\pm 0.121$} & 0.638 \scriptsize{$\pm 0.004$} & \textbf{0.844 \scriptsize{$\pm 0.005$}} \\
 & 3-mer Corr. $\uparrow$ & 0.591 \scriptsize{$\pm 0.059$} & 0.682 \scriptsize{$\pm 0.071$} & \textbf{0.969 \scriptsize{$\pm 0.013$}} & -0.354 \scriptsize{$\pm 0.007$} & 0.388 \scriptsize{$\pm 0.043$} & 0.725 \scriptsize{$\pm 0.011$} & 0.896 \scriptsize{$\pm 0.003$} \\
 & Shannon Div. $\uparrow$ & 0.487 \scriptsize{$\pm 0.302$} & 1.063 \scriptsize{$\pm 0.170$} & 1.980 \scriptsize{$\pm 0.002$} & 1.958 \scriptsize{$\pm 0.003$} & 1.851 \scriptsize{$\pm 0.019$} & \textbf{1.982 \scriptsize{$\pm 0.001$}} & 1.980 \scriptsize{$\pm 0.001$} \\
\midrule
\multirow{4}{*}{\textbf{SK-N-SH}}
 & MinGap Eval $\uparrow$ & 0.252 \scriptsize{$\pm 0.161$} & 1.170 \scriptsize{$\pm 0.715$} & -0.111 \scriptsize{$\pm 0.014$} & -0.082 \scriptsize{$\pm 0.040$} & 3.988 \scriptsize{$\pm 0.856$} & 3.202 \scriptsize{$\pm 0.258$} & \textbf{5.479 \scriptsize{$\pm 0.057$}} \\
 & Motif Corr. $\uparrow$ & 0.454 \scriptsize{$\pm 0.072$} & 0.435 \scriptsize{$\pm 0.008$} & 0.766 \scriptsize{$\pm 0.022$} & -0.001 \scriptsize{$\pm 0.009$} & 0.580 \scriptsize{$\pm 0.042$} & 0.502 \scriptsize{$\pm 0.053$} & \textbf{0.809 \scriptsize{$\pm 0.010$}} \\
 & 3-mer Corr. $\uparrow$ & 0.681 \scriptsize{$\pm 0.087$} & 0.658 \scriptsize{$\pm 0.135$} & \textbf{0.932 \scriptsize{$\pm 0.017$}} & -0.382 \scriptsize{$\pm 0.001$} & 0.431 \scriptsize{$\pm 0.074$} & 0.387 \scriptsize{$\pm 0.080$} & 0.912 \scriptsize{$\pm 0.001$} \\
 & Shannon Div. $\uparrow$ & 1.029 \scriptsize{$\pm 0.091$} & 1.019 \scriptsize{$\pm 0.327$} & 1.969 \scriptsize{$\pm 0.004$} & 1.826 \scriptsize{$\pm 0.001$} & 1.847 \scriptsize{$\pm 0.066$} & \textbf{1.982 \scriptsize{$\pm 0.001$}} & 1.972 \scriptsize{$\pm 0.001$} \\
\bottomrule
\end{tabular}%
}
\vspace{-15pt}
\end{table}

\textbf{Results.} Benchmarking results (Table~\ref{tab:enhancer_results}) demonstrate that DNA-CRAFT achieves a favorable trade-off between predicted cell-type specificity and biological fidelity. Across all three cell lines, it combines high differential activity with strong motif and 3-mer correlations. D3 achieves the highest 3-mer correlations and the lowest NTv2-based FBD scores (Table~\ref{tab:human_enhancer_extended}), indicating close agreement with the reference sequence distribution, but its near-zero or negative MinGap scores indicate poor steering toward cell-type specificity. Ctrl-DNA achieves higher MinGap than DNA-CRAFT on HepG2 and K562, while Ledidi also achieves higher MinGap than DNA-CRAFT on K562. DNA-CRAFT maintains higher motif correlations than either method. On SK-N-SH, DNA-CRAFT achieves the highest MinGap ($5.479$) and motif correlation ($0.809$), and retains the highest specificity under the alternative evaluation metrics LogSumExp and Mean Difference. Figure~\ref{fig:tradeoff_benchmarks} in the appendix visualizes the specificity-naturalness trade-off.

\subsection{Designing Immune Cell-Type-Specific Sequences}
\label{sec:immune_benchmark}

We further evaluated DNA-CRAFT's ability to design regulatory DNA for another data type (chromatin accessibility) and for cell types that are similar to each other (eight types of immune cells: CD4$^+$ T cells, CD8$^+$ T cells, Macrophages, Mast cells, Memory B cells, Natural Killer (NK) T cells, Naive T cells, and Plasma B cells). 

\textbf{Experimental Setup.} We utilized an Enformer model fine-tuned on single-cell ATAC-seq profiles from the CATlas dataset \citep{zhang_single-cell_2021, lal_grelu_2025} as our sequence-to-activity model. As in Section~\ref{sec:human_cell_line}, we trained separate Design and Evaluation models on non-overlapping homology clusters to prevent data leakage between optimization and evaluation. All methods generate sequences using the Design oracle. The final MinGap scores are reported using the held-out Evaluation oracle. We performed eight separate design experiments, configuring the objective to target one cell type while penalizing activity in the remaining seven. Evaluation followed the metrics defined in Section \ref{sec:human_cell_line}. Natural differentially accessible regions specific to each target cell type were used as references to calculate motif and 3-mer correlations. We benchmarked DNA-CRAFT (conditioned on ``Enhancer'', $\gamma=1.0$, lookahead size $B = 800$) against the same methods as the previous benchmark, except D3 and DRAKES, due to the computational cost of fine-tuning/training these methods on each task.

\begin{table}[h]
\centering
\caption{Comparison of methods for cell-type-specific regulatory sequence design across eight immune cell types. Shown are the MinGap specificity scores (evaluated under a held-out Evaluation oracle), Motif Spearman correlations, 3-mer Pearson correlations, and Shannon diversity. Values are reported as mean $\pm$ std over three independent runs and the best result is marked in \textbf{bold}.}
\label{tab:immune_enhancer_results}
\resizebox{\textwidth}{!}{%
\begin{tabular}{llccccc}
\toprule
\textbf{Target Cell Type} & \textbf{Metric} & \textbf{SMC} & \textbf{TDS} & \textbf{Ctrl-DNA} & \textbf{Ledidi} & \textbf{DNA-CRAFT (Ours)} \\
\midrule
\multirow{4}{*}{\textbf{CD4$^+$ T Cell}} & MinGap $\uparrow$ & -0.065 \scriptsize{$\pm 0.069$} & -0.016 \scriptsize{$\pm 0.002$} & -0.047 \scriptsize{$\pm 0.005$} & -0.105 \scriptsize{$\pm 0.006$} & \textbf{0.075} \scriptsize{$\pm 0.001$} \\
 & Motif Corr. $\uparrow$ & 0.478 \scriptsize{$\pm 0.048$} & 0.363 \scriptsize{$\pm 0.045$} & 0.261 \scriptsize{$\pm 0.031$} & 0.094 \scriptsize{$\pm 0.155$} & \textbf{0.616} \scriptsize{$\pm 0.024$} \\
 & 3-mer Corr. $\uparrow$ & \textbf{0.811} \scriptsize{$\pm 0.056$} & 0.802 \scriptsize{$\pm 0.073$} & 0.461 \scriptsize{$\pm 0.035$} & 0.479 \scriptsize{$\pm 0.039$} & 0.766 \scriptsize{$\pm 0.007$} \\
 & Shannon Div. $\uparrow$ & 1.227 \scriptsize{$\pm 0.160$} & 1.339 \scriptsize{$\pm 0.285$} & 1.779 \scriptsize{$\pm 0.022$} & \textbf{1.983} \scriptsize{$\pm 0.001$} & 1.919 \scriptsize{$\pm 0.002$} \\
\midrule
\multirow{4}{*}{\textbf{CD8$^+$ T Cell}} & MinGap $\uparrow$ & -0.022 \scriptsize{$\pm 0.019$} & -0.022 \scriptsize{$\pm 0.018$} & \textbf{0.084} \scriptsize{$\pm 0.004$} & -0.048 \scriptsize{$\pm 0.003$} & 0.059 \scriptsize{$\pm 0.008$} \\
 & Motif Corr. $\uparrow$ & 0.443 \scriptsize{$\pm 0.158$} & 0.360 \scriptsize{$\pm 0.302$} & 0.076 \scriptsize{$\pm 0.054$} & 0.084 \scriptsize{$\pm 0.215$} & \textbf{0.680} \scriptsize{$\pm 0.055$} \\
 & 3-mer Corr. $\uparrow$ & \textbf{0.726} \scriptsize{$\pm 0.115$} & 0.716 \scriptsize{$\pm 0.137$} & 0.221 \scriptsize{$\pm 0.010$} & 0.045 \scriptsize{$\pm 0.140$} & 0.558 \scriptsize{$\pm 0.010$} \\
 & Shannon Div. $\uparrow$ & 1.088 \scriptsize{$\pm 0.282$} & 1.400 \scriptsize{$\pm 0.097$} & 1.684 \scriptsize{$\pm 0.018$} & \textbf{1.983} \scriptsize{$\pm 0.001$} & 1.918 \scriptsize{$\pm 0.002$} \\
\midrule
\multirow{4}{*}{\textbf{Macrophage}} & MinGap $\uparrow$ & 0.056 \scriptsize{$\pm 0.027$} & 0.053 \scriptsize{$\pm 0.036$} & 0.423 \scriptsize{$\pm 0.123$} & 0.072 \scriptsize{$\pm 0.013$} & \textbf{0.947} \scriptsize{$\pm 0.001$} \\
 & Motif Corr. $\uparrow$ & 0.498 \scriptsize{$\pm 0.048$} & 0.447 \scriptsize{$\pm 0.079$} & 0.591 \scriptsize{$\pm 0.088$} & 0.147 \scriptsize{$\pm 0.154$} & \textbf{0.827} \scriptsize{$\pm 0.004$} \\
 & 3-mer Corr. $\uparrow$ & 0.660 \scriptsize{$\pm 0.207$} & 0.825 \scriptsize{$\pm 0.027$} & 0.430 \scriptsize{$\pm 0.050$} & 0.079 \scriptsize{$\pm 0.150$} & \textbf{0.898} \scriptsize{$\pm 0.003$} \\
 & Shannon Div. $\uparrow$ & 1.275 \scriptsize{$\pm 0.125$} & 1.429 \scriptsize{$\pm 0.076$} & 1.738 \scriptsize{$\pm 0.005$} & \textbf{1.983} \scriptsize{$\pm 0.001$} & 1.971 \scriptsize{$\pm 0.000$} \\
\midrule
\multirow{4}{*}{\textbf{Mast Cell}} & MinGap $\uparrow$ & -0.047 \scriptsize{$\pm 0.038$} & -0.021 \scriptsize{$\pm 0.012$} & 0.852 \scriptsize{$\pm 0.099$} & -0.056 \scriptsize{$\pm 0.007$} & \textbf{0.886} \scriptsize{$\pm 0.002$} \\
 & Motif Corr. $\uparrow$ & 0.474 \scriptsize{$\pm 0.082$} & 0.540 \scriptsize{$\pm 0.080$} & 0.444 \scriptsize{$\pm 0.070$} & 0.113 \scriptsize{$\pm 0.182$} & \textbf{0.881} \scriptsize{$\pm 0.004$} \\
 & 3-mer Corr. $\uparrow$ & 0.800 \scriptsize{$\pm 0.080$} & 0.856 \scriptsize{$\pm 0.000$} & 0.395 \scriptsize{$\pm 0.123$} & 0.085 \scriptsize{$\pm 0.150$} & \textbf{0.912} \scriptsize{$\pm 0.006$} \\
 & Shannon Div. $\uparrow$ & 1.482 \scriptsize{$\pm 0.012$} & 1.397 \scriptsize{$\pm 0.179$} & 1.835 \scriptsize{$\pm 0.086$} & \textbf{1.983} \scriptsize{$\pm 0.001$} & 1.982 \scriptsize{$\pm 0.001$} \\
\midrule
\multirow{4}{*}{\textbf{Memory B Cell}} & MinGap $\uparrow$ & -0.042 \scriptsize{$\pm 0.030$} & -0.055 \scriptsize{$\pm 0.055$} & -0.044 \scriptsize{$\pm 0.007$} & -0.151 \scriptsize{$\pm 0.005$} & \textbf{0.467} \scriptsize{$\pm 0.021$} \\
 & Motif Corr. $\uparrow$ & 0.392 \scriptsize{$\pm 0.026$} & 0.591 \scriptsize{$\pm 0.038$} & 0.287 \scriptsize{$\pm 0.203$} & 0.076 \scriptsize{$\pm 0.181$} & \textbf{0.860} \scriptsize{$\pm 0.003$} \\
 & 3-mer Corr. $\uparrow$ & 0.785 \scriptsize{$\pm 0.048$} & 0.798 \scriptsize{$\pm 0.023$} & 0.659 \scriptsize{$\pm 0.147$} & 0.063 \scriptsize{$\pm 0.139$} & \textbf{0.937} \scriptsize{$\pm 0.006$} \\
 & Shannon Div. $\uparrow$ & 1.236 \scriptsize{$\pm 0.264$} & 1.330 \scriptsize{$\pm 0.187$} & 1.950 \scriptsize{$\pm 0.008$} & \textbf{1.983} \scriptsize{$\pm 0.001$} & 1.982 \scriptsize{$\pm 0.002$} \\
\midrule
\multirow{4}{*}{\textbf{NK T Cell}} & MinGap $\uparrow$ & -0.039 \scriptsize{$\pm 0.023$} & -0.016 \scriptsize{$\pm 0.005$} & 0.130 \scriptsize{$\pm 0.039$} & -0.122 \scriptsize{$\pm 0.005$} & \textbf{0.344} \scriptsize{$\pm 0.006$} \\
 & Motif Corr. $\uparrow$ & 0.384 \scriptsize{$\pm 0.124$} & 0.353 \scriptsize{$\pm 0.271$} & 0.315 \scriptsize{$\pm 0.026$} & 0.078 \scriptsize{$\pm 0.202$} & \textbf{0.804} \scriptsize{$\pm 0.006$} \\
 & 3-mer Corr. $\uparrow$ & 0.780 \scriptsize{$\pm 0.065$} & 0.742 \scriptsize{$\pm 0.121$} & 0.455 \scriptsize{$\pm 0.037$} & 0.043 \scriptsize{$\pm 0.161$} & \textbf{0.784} \scriptsize{$\pm 0.013$} \\
 & Shannon Div. $\uparrow$ & 1.431 \scriptsize{$\pm 0.033$} & 1.401 \scriptsize{$\pm 0.096$} & 1.871 \scriptsize{$\pm 0.023$} & \textbf{1.983} \scriptsize{$\pm 0.001$} & 1.963 \scriptsize{$\pm 0.002$} \\
\midrule
\multirow{4}{*}{\textbf{Naive T Cell}} & MinGap $\uparrow$ & -0.031 \scriptsize{$\pm 0.017$} & -0.064 \scriptsize{$\pm 0.037$} & 0.229 \scriptsize{$\pm 0.031$} & -0.126 \scriptsize{$\pm 0.007$} & \textbf{0.610} \scriptsize{$\pm 0.003$} \\
 & Motif Corr. $\uparrow$ & 0.422 \scriptsize{$\pm 0.123$} & 0.410 \scriptsize{$\pm 0.080$} & 0.644 \scriptsize{$\pm 0.052$} & 0.105 \scriptsize{$\pm 0.154$} & \textbf{0.859} \scriptsize{$\pm 0.006$} \\
 & 3-mer Corr. $\uparrow$ & 0.618 \scriptsize{$\pm 0.145$} & 0.769 \scriptsize{$\pm 0.085$} & 0.684 \scriptsize{$\pm 0.038$} & 0.061 \scriptsize{$\pm 0.128$} & \textbf{0.942} \scriptsize{$\pm 0.004$} \\
 & Shannon Div. $\uparrow$ & 1.413 \scriptsize{$\pm 0.103$} & 1.361 \scriptsize{$\pm 0.230$} & 1.906 \scriptsize{$\pm 0.040$} & \textbf{1.983} \scriptsize{$\pm 0.001$} & 1.981 \scriptsize{$\pm 0.001$} \\
\midrule
\multirow{4}{*}{\textbf{Plasma B Cell}} & MinGap $\uparrow$ & -0.019 \scriptsize{$\pm 0.014$} & -0.037 \scriptsize{$\pm 0.019$} & 0.458 \scriptsize{$\pm 0.063$} & -0.081 \scriptsize{$\pm 0.001$} & \textbf{0.785} \scriptsize{$\pm 0.007$} \\
 & Motif Corr. $\uparrow$ & 0.454 \scriptsize{$\pm 0.072$} & 0.348 \scriptsize{$\pm 0.277$} & 0.482 \scriptsize{$\pm 0.057$} & 0.024 \scriptsize{$\pm 0.165$} & \textbf{0.827} \scriptsize{$\pm 0.006$} \\
 & 3-mer Corr. $\uparrow$ & 0.800 \scriptsize{$\pm 0.034$} & 0.643 \scriptsize{$\pm 0.038$} & 0.433 \scriptsize{$\pm 0.033$} & 0.042 \scriptsize{$\pm 0.167$} & \textbf{0.879} \scriptsize{$\pm 0.010$} \\
 & Shannon Div. $\uparrow$ & 1.310 \scriptsize{$\pm 0.098$} & 1.138 \scriptsize{$\pm 0.268$} & 1.806 \scriptsize{$\pm 0.071$} & \textbf{1.983} \scriptsize{$\pm 0.001$} & 1.969 \scriptsize{$\pm 0.001$} \\
\bottomrule
\end{tabular}%
}
\end{table}

\textbf{Results.} Decoupling chromatin accessibility among closely related cell types (e.g., CD4$^+$ versus CD8$^+$ T cells) is a challenging generative task. As shown in Table \ref{tab:immune_enhancer_results}, DNA-CRAFT is the only method that achieves positive predicted differential accessibility (MinGap) across all eight cell types, while maintaining high motif correlations. Ctrl-DNA achieves positive MinGap for several cell types, but has lower motif correlations than DNA-CRAFT, particularly for CD4$^+$ and CD8$^+$ T cells. We further evaluated DNA-CRAFT on eight brain cell types from the CATlas dataset, with the experimental setup and complete results provided in Appendix~\ref{app:brain_benchmark}.

\subsection{Ablation Studies}
\label{sec:ablations}

To isolate the impact of our algorithmic components, we conducted ablation studies using the SK-N-SH cell line enhancer design task. We generated 128 sequences per configuration and evaluated the trade-offs between computational efficiency (wall time), differential activity, and realism across 3 independent runs. We selected SK-N-SH for ablations because the highest mean MinGap attained by any evaluated method was lower than for HepG2 or K562 (Table~\ref{tab:enhancer_results}), motivating its use as a challenging test case.

\begin{figure}[h]
    \vspace{-5pt}
    \centering
    \includegraphics[width=0.9\textwidth]{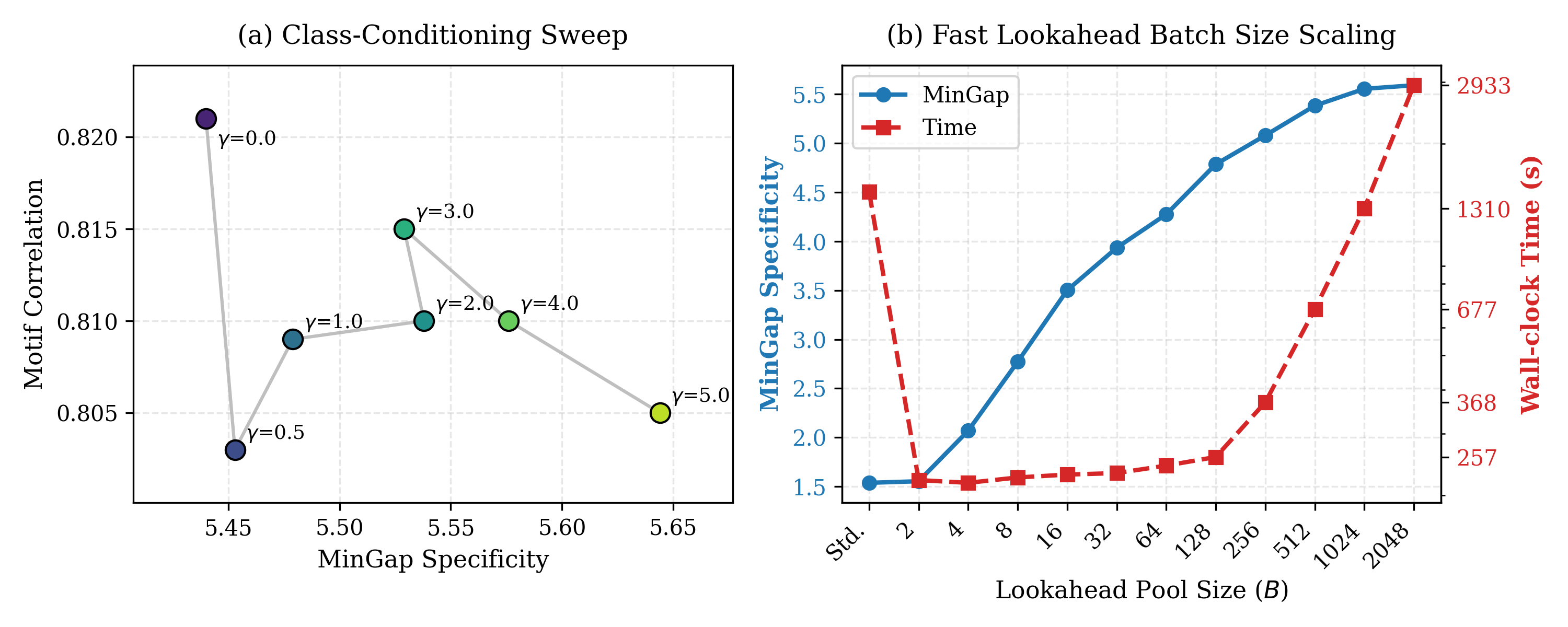}
    \caption{Ablation of DNA-CRAFT class-conditioned expansion and lookahead mechanisms for SK-N-SH enhancer design. \textbf{(a)} Class-conditioning sweep across guidance scales ($\gamma$) with MinGap specificity and motif correlation reported. \textbf{(b)}  Fast lookahead pruning batch size ($B$) scaling compared to standard full rollouts (Std.), with MinGap specificity and compute wall time reported.}
    \label{fig:ablation_sweep}
    \vspace{-10pt}
\end{figure}

\textbf{Class conditioning improves specificity while retaining motif fidelity.} We evaluated class-conditioned tree expansion by sweeping the CFG scale $\gamma\in\{0.0,0.5,1.0,2.0,3.0,4.0,5.0\}$. At $\gamma=0$, sampling uses the null conditioning label of the same network. Mean MinGap generally increases across the sweep, from $5.440$ at $\gamma=0$ to $5.644$ at $\gamma=5$, with a small decrease between $\gamma=2$ and $\gamma=3$ (Figure~\ref{fig:ablation_sweep}a). Motif correlation remains between $0.803$ and $0.821$ without a monotonic trend. These results show improved predicted specificity at higher guidance scales while retaining similar motif-frequency correlations.

\textbf{Fast lookahead pruning increases specificity while reducing compute cost.} To quantify the computational and performance benefits of the fast lookahead pruning, we swept the lookahead batch size $B \in \{2, 4, 8, 16, 32, 64, 128, 256, 512, 1024, 2048\}$ with a retained-child limit of $M=8$, giving $\min(B,8)$ retained children per expansion and compared it to standard full rollouts (Std.). Our results (Figure \ref{fig:ablation_sweep}b) demonstrate that as pool size $B$ scales, the mean MinGap specificity improves monotonically from $1.554$ to $5.590$, while motif and 3-mer correlations vary. The one-step lookahead configurations achieve higher mean specificity than the full-rollout baseline, although their candidate budgets differ. The full-rollout baseline retains higher motif and 3-mer correlations, although these differences are modest at $B=256$: $0.858$ versus $0.828$ for motif correlation and $0.947$ versus $0.924$ for 3-mer correlation (Table~\ref{tab:lookahead_sweep_full}). Crucially, even with a massive candidate pool of $B=256$, the method achieves a high MinGap of $5.079$ in just $368$ seconds. In contrast, standard MCTS ($M=B$) relying on full iterative rollouts achieves a much lower MinGap score of $1.538$ at four times the compute time of $1459$ seconds. This shows that our fast lookaheads improve both the specificity and the computational efficiency of the tree search. 

\begin{table}[h]
    \vspace{-5pt}
    \centering
    \caption{Evaluating different specificity reward formulations during the Monte Carlo Tree Guidance search ($M=8$, $B=800$) for SK-N-SH enhancer design. Values are reported as mean $\pm$ standard deviation over 3 independent runs and the best result is marked in \textbf{bold}.}
    \label{tab:reward_ablation}
    \resizebox{1.0\textwidth}{!}{
    \begin{tabular}{lcccc}
        \toprule
        \textbf{Reward Formulation} & \textbf{MinGap} ($\uparrow$) & \textbf{Mean Diff.} ($\uparrow$) & \textbf{Motif Corr.} ($\uparrow$) & \textbf{3-mer Corr.} ($\uparrow$) \\
        \midrule
        Pareto Reward & 0.495 \scriptsize{$\pm 0.059$} & 0.643 \scriptsize{$\pm 0.068$} & 0.810 \scriptsize{$\pm 0.028$} & \textbf{0.941} \scriptsize{$\pm 0.010$} \\
        Mean activity difference    & 4.551 \scriptsize{$\pm 0.035$} & 6.124 \scriptsize{$\pm 0.023$} & \textbf{0.814} \scriptsize{$\pm 0.009$} & 0.919 \scriptsize{$\pm 0.000$} \\
        MinGap & \textbf{5.531} \scriptsize{$\pm 0.056$} & 5.865 \scriptsize{$\pm 0.050$} & 0.801 \scriptsize{$\pm 0.019$} & 0.891 \scriptsize{$\pm 0.006$} \\
        
        MinGap LogSumExp & 5.479 \scriptsize{$\pm 0.057$} & \textbf{6.134} \scriptsize{$\pm 0.035$} & 0.809 \scriptsize{$\pm 0.010$} & 0.912 \scriptsize{$\pm 0.001$} \\
        \midrule
    \end{tabular}%
    }
    \vspace{-5pt}
\end{table}

\textbf{Specificity reward formulation.} We compared four reward formulations within the tree search: Pareto dominance, mean activity difference, hard MinGap, and our $\mathrm{LogSumExp}$ MinGap with a relative archive (Table~\ref{tab:reward_ablation}). The Pareto reward achieves strong motif correlation ($0.810$) and the highest 3-mer correlation ($0.941$), while yielding lower cell-type specificity (MinGap $0.495$). All three scalar specificity rewards achieve higher mean MinGap, highlighting the benefit of directly optimizing the separation between target and off-target activity. Among scalar formulations, mean activity difference achieves a MinGap of $4.551$, while hard MinGap achieves the highest specificity ($5.531$) with lower mean differential activity ($5.865$) and 3-mer correlation ($0.891$). Our $\mathrm{LogSumExp}$ MinGap with rank-based archive normalization (Equation~\ref{eq:rank_reward}) achieves a favorable balance, with comparable specificity ($5.479$), the highest reported mean differential activity ($6.134$), and strong naturalness (motif correlation $0.809$, 3-mer correlation $0.912$). These results illustrate the trade-off between worst-case specificity, mean differential activity, and the measured naturalness metrics.

\section{Conclusion}
\label{sec:conclusion}
We introduced DNA-CRAFT, a method for designing regulatory DNA with high predicted cell-type-specific activity while preserving naturalness. DNA-CRAFT formulates sequence design as inference-time search over the denoising trajectory of a frozen discrete diffusion model, combining class-conditioned expansion, fast lookahead pruning, and a MinGap specificity reward. Across benchmarks spanning human cell lines and primary immune cell types, DNA-CRAFT achieves a favorable trade-off between predicted specificity and naturalness metrics. Ablation studies delineate how the search components and reward formulation influence specificity, naturalness, and computational cost. Limitations include dependence on sequence-to-activity models and evaluation against few off-target cell types. The designs have not been experimentally validated. Future directions include integrating larger pre-trained DNA foundation models as the generative backbone, extending the framework to additional tissues and species, and experimentally validating designs for clinically relevant cell types.

\begin{ack}
\textbf{Broader Impact.} Methods for designing regulatory DNA may support basic research, cell engineering, and therapeutic development. Their outputs remain subject to prediction errors and may exhibit unintended activity outside the evaluated cell types. Experimental validation and appropriate oversight are therefore necessary before downstream use.

\textbf{Acknowledgements.} We thank the members of the Bock Lab and the AI Institute of the Medical University of Vienna for their help and advice. We also thank Albert López i Serrano for helpful discussions. 
 
\textbf{Funding.} C.B. is supported by a European Research Council (ERC) Consolidator Grant (101001971). Computational resources were provided by CeMM with infrastructure supported by CeMM’s IT Services team.

\textbf{Competing interests.} C.B. is a cofounder and scientific advisor of Myllia Biotechnology and Neurolentech. The remaining authors declare no competing interests.
\end{ack}

\bibliographystyle{plainnat}
\bibliography{neurips2026_conference}
\newpage

\appendix
\section{Appendix}
The appendix provides related work (Appendix~\ref{app:related_work}), the complete DNA-CRAFT algorithm (Appendix~\ref{app:complete_algorithm}), and diffusion model training details and test-set evaluations (Appendix~\ref{app:diffusion_training}). Appendix~\ref{app:class_cond_details} describes class-conditioned sampling and motif analysis. Appendix~\ref{app:benchmark_details} provides benchmark implementation details, extended evaluations including eight brain cell types, and training and validation results for the sequence-to-activity models. Appendix~\ref{app:extended_ablations} reports the full ablation results and cross-method evaluation under alternative specificity metrics. Finally, Appendices~\ref{app:tree_vs_greedy}, \ref{app:tree_depth}, and~\ref{app:map_validation} present the comparison with a tree-free greedy baseline, tree depth statistics, and evaluation of the one-step MAP estimator, respectively.

\subsection{Related Work}

\label{app:related_work}
Machine learning methods for regulatory element design rely on sequence-to-activity neural networks that predict gene expression or chromatin accessibility from DNA sequences \citep{enformer, linder_predicting_2025}. By generalizing to unseen sequences, these models serve as differentiable sequence-to-activity models that enable computational design of regulatory elements with desired activity profiles. Existing approaches generally fall into two categories: sequence optimization methods and generative model-based methods.

\textbf{Sequence Optimization Methods. }These methods treat regulatory DNA design as a search problem over a combinatorial sequence space, using a pre-trained sequence-to-activity model as the objective function. Classical approaches, including simulated annealing \citep{laarhoven_simulated_1987}, \textit{in silico} directed evolution, AdaLead \citep{sinai2020adalead}, and gradient-based editing \citep{schreiber_programmatic_2025, linder2021fast}, start from random or known sequences and iteratively mutate them to maximize predicted activity. More recently, reinforcement learning techniques such as DyNA-PPO \citep{angermueller_model-based_2019} and GFlowNets \citep{GFlowNet} have been applied to navigate this search space more effectively. Because these methods optimize a scalar reward without an explicit distributional constraint, they are prone to generating sequences that exploit biases in the prediction model and diverge from the regulatory grammar of natural DNA \citep{vaishnav2022evolution}.

\textbf{Generative Models for Regulatory DNA.} Generative approaches learn the underlying distribution of natural DNA sequences and use this learned prior to constrain the design space. Recent methods employ autoregressive genomic language models \citep{nguyen2023hyenadna, schiff_caduceus_2024, reglm} and discrete diffusion models \citep{avdeyev2023dirichlet, sarkar_designing_2025, dasilva_designing_2026} to capture long-range sequence dependencies with high fidelity. To steer such generative priors toward cell-type-specific regulatory activity, three guidance paradigms have emerged. \textit{Classifier-free guidance} (CFG) builds class-conditioned diffusion models that can be steered toward desired sequence classes at sampling time \citep{schiff_simple_2025}. While CFG preserves sequence quality, conditioning on broad regulatory classes alone does not provide sufficient specificity for fine-grained cell-type discrimination. \textit{RL-based fine-tuning} updates the generative model weights to maximize a reward function. Ctrl-DNA combines genomic language models with constrained RL for cell-type-specific design \citep{chen_ctrl-dna_2025}, and DRAKES backpropagates reward gradients through a discrete diffusion model via the Gumbel-Softmax trick \citep{wang_fine-tuning_2025}. While effective, fine-tuning is computationally expensive and must be repeated for every new design objective. \textit{Inference-time alignment} methods steer the sampling process of a frozen generative model using external guidance signals, avoiding the cost of retraining \citep{uehara2025inferencetimealignmentdiffusionmodels}. Techniques such as soft value-based decoding \citep{li_derivative-free_2024}, sequential Monte Carlo (SMC) \citep{phillips_particle_2024}, twisted diffusion sampling (TDS) \citep{wu_practical_2024}, and classifier-based guidance \citep{nisonoff_unlocking_2025} use reward-weighted resampling or auxiliary gradients to steer the generative process. However, these methods typically optimize for high activity in a single cell type without explicitly suppressing off-target activity, limiting their applicability to strict cell-type specificity tasks.

\textbf{Search-Based Guidance for Diffusion Models.} An alternative to gradient-based or resampling-based inference-time alignment is to formulate the reverse diffusion process as a sequential decision problem and apply tree search algorithms to guide generation. Monte Carlo tree search (MCTS) has been adapted to continuous diffusion models as a planner \citep{yoon2026montecarlotreediffusion, yoon2025fastmontecarlotree}, and extended to discrete diffusion for multi-objective peptide design \citep{tang_peptune_2024}. Multi-objective guided discrete flow matching has also been explored in related biological design tasks \citep{chen2025multiobjectiveguideddiscreteflowmatching}. These methods use tree search to balance reward maximization with distributional fidelity by evaluating candidate completions of partially denoised states.

\subsection{Complete Algorithm}
\label{app:complete_algorithm}
We provide the complete DNA-CRAFT algorithm in Algorithm~\ref{alg:dna_craft}. The procedure implements class-conditioned Monte Carlo tree diffusion with fast lookahead pruning, as described in Section~\ref{sec:methods}. Candidates are ranked using the smooth MinGap score $g$ (Equation~\ref{eq:smooth_mingap}), while the archive-relative rank rewards $\rho_m$ (Equation~\ref{eq:rank_reward}) are used for tree updates. We use $N(\bm{z})$ to denote the visit count and $\bar{Q}(\bm{z})$ to denote the accumulated backpropagated rank reward divided by the visit count of node $\bm{z}$. Each expansion adds the sum of the retained children's rank rewards to the accumulated reward of each ancestor and increments its visit count once. The soft-optimal policy in Equation~\ref{eq:optimal_policy} motivates the search, but DNA-CRAFT uses finite candidate sampling, lookahead pruning, and UCT selection rather than sampling exactly from this policy. Consequently, $\lambda$ is not an explicit parameter of Algorithm~\ref{alg:dna_craft}.

\begin{algorithm}
\caption{DNA-CRAFT: Class-Conditioned Monte Carlo Tree Diffusion}
\label{alg:dna_craft}
\begin{algorithmic}[1]
\State \textbf{Input:} Denoiser, Design model $\bm{s}$, regulatory class $\bm{y}$, guidance scale $\gamma$
\State \phantom{\textbf{Input:}} Proposal count $B$, child limit $M$, coefficient $c$, iteration limit $N_{\text{iter}}$
\State \phantom{\textbf{Input:}} Archive capacity $N_{\text{max}}$
\State Initialize a fully masked root with $N=1$ and $\bar{Q}=0$
\State Initialize archive $\mathcal{G}^*\gets\emptyset$
\For{$k=1,\ldots,N_{\text{iter}}$}
    \State $\text{node}\gets\text{root}$
    \While{$\text{node}$ has children and is not terminal}
        \State $\text{node}\gets\displaystyle\operatorname*{arg\,max}_{\bm{z}\in\operatorname{children}(\text{node})}\left[\bar{Q}(\bm{z})+c\sqrt{\frac{\log N(\text{node})}{N(\bm{z})}}\right]$
    \EndWhile
    \If{$\text{node}$ is terminal}
        \State Skip the remainder of this iteration
    \Else
        \State Let $\bm{z}_t\gets\text{node}$ and let $s<t$ be the next diffusion time
        \For{$b=1,\ldots,B$}
            \State Sample $\bm{z}_{s,b}$ using Equation~\ref{eq:child_sampling} and retain revealed bases
            \State Obtain $\hat{\bm{x}}_{0,b}$ using Equation~\ref{eq:x0_map}
            \State $g_b\gets g(\hat{\bm{x}}_{0,b})$ using Equation~\ref{eq:smooth_mingap}
        \EndFor
        \State $\mathcal{I}_M\gets$ indices of the $\min(M,B)$ highest-scoring candidates, in decreasing order
        \For{$m\in\mathcal{I}_M$}
            \State Compute $\rho_m$ against the current archive using Equation~\ref{eq:rank_reward}
            \If{$\mathcal{G}^*=\emptyset$}
                \State Insert $\hat{\bm{x}}_{0,m}$ into $\mathcal{G}^*$
            \ElsIf{$g_m\geq\max_{\bm{x}^*\in\mathcal{G}^*}g(\bm{x}^*)$}
                \State Insert or update $\hat{\bm{x}}_{0,m}$ in $\mathcal{G}^*$
                \State Enforce capacity $N_{\text{max}}$ using the archive rule described in Section~\ref{sec:methods}
            \EndIf
            \State Add $\bm{z}_{s,m}$ as a child of $\bm{z}_t$, with $N(\bm{z}_{s,m})=1$ and $\bar{Q}(\bm{z}_{s,m})=\rho_m$
        \EndFor
        \For{each node $\bm{z}$ from $\bm{z}_t$ to the root}
            \State $\displaystyle\bar{Q}(\bm{z})\gets\frac{N(\bm{z})\bar{Q}(\bm{z})+\sum_{m\in\mathcal{I}_M}\rho_m}{N(\bm{z})+1}$
            \State $N(\bm{z})\gets N(\bm{z})+1$
        \EndFor
    \EndIf
\EndFor
\State \textbf{Output:} Archive $\mathcal{G}^*$ of admitted record-scoring completions
\end{algorithmic}
\end{algorithm}

\subsection{DNA-CRAFT Diffusion Model Training Details}
\label{app:diffusion_training}

\textbf{Forward noising process.} The forward process progressively corrupts a naturally occurring DNA sequence \(\bm{x}\) into a sequence of discrete variables \(\bm{z}_t\) over continuous time \(t \in [0, 1]\), where unmasked tokens (base pairs) transition to the masked token \(\bm{m}\). The marginal probability of a latent state \(\bm{z}_t\) given the input DNA sequence \(\bm{x}\) is defined as a categorical distribution \(  
q(\bm{z}_t | \bm{x})=\mathrm{Cat}(\bm{z}_t;\alpha_t \bm{x}+(1-\alpha_t)\bm{m})
\), where $\alpha_t \in [0, 1]$ is a monotonically decreasing noise schedule. Concretely, this means that at time $t$, each token independently remains as the original nucleotide with probability $\alpha_t$ or is replaced with probability $1 - \alpha_t$ by the absorbing state mask token $\bm{m}$.

\textbf{Reverse denoising process.} The reverse process reconstructs the sequence by progressively unmasking positions from a fully masked sequence. Conditioned on the original DNA sequence \(\bm{x}\), the time-reversal process for time steps \(s < t\) is
\begin{equation}
\label{eq:time_reversal}
\resizebox{0.5\linewidth}{!}{$%
q(\bm{z}_s | \bm{z}_t,\bm{x})=
\begin{cases}
\mathrm{Cat}(\bm{z}_s;\bm{z}_t), & \bm{z}_t\neq \bm{m},\\
\mathrm{Cat}\Bigl(\bm{z}_s;\dfrac{(1-\alpha_s)\bm{m}+(\alpha_s-\alpha_t)\bm{x}}{1-\alpha_t}\Bigr), & \bm{z}_t=\bm{m}.
\end{cases}
$}%
\end{equation}
We train a class-conditioned backbone neural network denoted as $\bm{x}_\theta(\bm{z}_t,t,\bm{y})$ to predict categorical probabilities over the clean nucleotides given the noisy input $\bm{z}_t$ and regulatory class $\bm{y}$. The learned reverse transition is obtained by substituting these predictions into Equation~\ref{eq:time_reversal} at each position, with a loss function that simplifies to a weighted average of masked language modeling losses. To generate a DNA sequence of length $L$, we ancestrally sample from the learned reverse process, starting with a completely masked sequence $\bm{z}_t^{1:L}$ at $t=1$. Conditional on the current sequence and regulatory class, the reverse transition factorizes across positions:
\begin{equation}
\label{eq:reverse_transition}
p_\theta(\bm{z}_s^{1:L}\mid\bm{z}_t^{1:L},\bm{y}) = \prod_{\ell=1}^{L}p_\theta(\bm{z}_s^\ell\mid\bm{z}_t^{1:L},\bm{y}).
\end{equation}

\textbf{Dataset Curation.} We trained DNA-CRAFT using the ENCODE Registry of candidate cis-Regulatory Elements V4 \citep{moore_expanded_2026}. To capture cross-species regulatory grammar and maximize biological diversity, we integrated data from both human (hg38) and mouse (mm10) genomes. The final dataset comprises 2,348,854 annotated regions across 1,888 human cell types and 926,843 annotated regions across 366 mouse cell types.

\textbf{Data Pre-Processing.} DNA sequences corresponding to the regulatory regions were extracted from their respective reference genomes. All sequences were centered and padded to a fixed length of 350 base pairs (bp) using a distinct \texttt{[PAD]} token. To accommodate variable effective lengths, we implemented a masking mechanism within both the model backbone and the diffusion loss function, ensuring that padding tokens are excluded from the generative process. We applied reverse complement augmentation by taking either the forward or reverse complement strand with equal probability during training.

\textbf{Class Consolidation.} The ENCODE V4 Registry categorizes regulatory elements based on biochemical signatures derived from DNase hypersensitivity, histone modifications, and CTCF binding. The original classification schema distinguishes between:
\begin{itemize}
    \item \textbf{Promoter-like (PLS):} Accessible regions proximal to transcription start sites (TSS) enriched for H3K4me3.
    \item \textbf{Enhancer-like (ELS):} Accessible regions enriched for H3K27ac, further stratified into proximal (pELS) and distal (dELS) based on TSS proximity.
    \item \textbf{CA-H3K4me3:} Accessible regions with H3K4me3 enrichment but low H3K27ac, often indicative of poised or primed regulatory states.
    \item \textbf{CA-CTCF:} Accessible regions enriched for CTCF binding sites with low histone acetylation, often indicating insulators.
    \item \textbf{CA-TF, CA, and TF:} Elements defined primarily by chromatin accessibility or transcription factor binding, lacking canonical histone marks.
\end{itemize}

We consolidated these fine-grained categories into five broad functional classes for DNA-CRAFT conditioning (Table \ref{tab:class_consolidation}).

\begin{table}[ht]
    \centering
    \caption{Mapping of ENCODE V4 cCRE classifications to DNA-CRAFT conditioning labels.}
    \label{tab:class_consolidation}
    \begin{tabular}{@{}ll@{}}
        \toprule
        \textbf{DNA-CRAFT Class} & \textbf{Original ENCODE Class} \\
        \midrule
        Promoter & PLS \\
        \addlinespace[0.4em]
        Enhancer & dELS, pELS \\
        \addlinespace[0.4em]
        CTCF & CA-CTCF \\
        \addlinespace[0.4em]
        Poised & CA-H3K4me3, TF \\
        \addlinespace[0.4em]
        Open Chromatin & CA, CA-TF \\
        \bottomrule
    \end{tabular}
\end{table}

\textbf{Cross-Species Train-Test-Validation Split.} To ensure generalization and prevent overfitting due to sequence homology, we constructed our data splits using a graph-based clustering approach adapted from the Enformer training protocol \citep{enformer}. 

We constructed an undirected graph $G=(V, E)$ in which the vertices $V$ represent approximately 3.2 million processed regulatory elements. Edges $E$ were defined to capture both orthology and sequence similarity:
\begin{enumerate}[noitemsep, topsep=2pt, leftmargin=*]
    \item \textbf{Homology Edges:} An edge connects a human cCRE and a mouse cCRE if they share a sequence alignment of $>100$ bp, determined via the hg38-mm10 syntenic nets \citep{kent_human_2002}.
    \item \textbf{Overlap Edges:} Within a single species, an edge connects any two cCREs that share sequence similarity of more than $100$ bp.
\end{enumerate}

We computed the connected components of $G$ to define independent sequence clusters. These clusters were randomly partitioned into training (90\%), validation (5\%), and test (5\%) sets, ensuring a strict separation of homologous sequences across splits. 

\textbf{Hyperparameters and Training.} We trained the class-conditioned diffusion model on the processed ENCODE dataset for 100 epochs using the AdamW optimizer and a global batch size of 4,096. During training, the regulatory class label was replaced with the null conditioning label with probability $0.1$ to support classifier-free guidance. Model hyperparameters are provided in Table~\ref{tab:hyperparams}.

\begin{table}[ht]
    \centering
    \caption{DNA-CRAFT Hyperparameters. The model utilizes a bidirectional DiMamba backbone.}
    \label{tab:hyperparams}
    \begin{tabular}{@{}ll@{}}
        \toprule
        \textbf{Hyperparameter} & \textbf{Value} \\
        \midrule
        \multicolumn{2}{l}{\textit{Backbone Architecture (DiMamba)}} \\
 Model Parameters&1.93 Million\\
        Sequence Length ($L$) & 350 \\
        Hidden Dimension ($d_{\text{model}}$) & 128 \\
        Mamba Blocks & 10 \\
        Bidirectional Strategy & Addition (Tied weights) \\
        Dropout & 0.1 \\
        \midrule
        \multicolumn{2}{l}{\textit{Diffusion Process}} \\
        Noise Schedule & Cosine \\
        Time Conditioning & True \\
        \midrule
        \multicolumn{2}{l}{\textit{Optimization}} \\
        Optimizer & AdamW \\
        Learning Rate & $1 \times 10^{-3}$ \\
        Training Epochs & 100 \\
        Batch Size & 4,096 \\
        \midrule
        \multicolumn{2}{l}{\textit{Conditioning (CFG)}} \\
        Condition Type & Regulatory Element Class \\
        Number of Classes & 5 \\
        Condition Dropout & 0.1 \\
        Conditioning Dim & 128 \\
        \bottomrule
    \end{tabular}
\end{table}

\begin{figure}[ht] 
    \centering 
    \includegraphics[width=\linewidth]{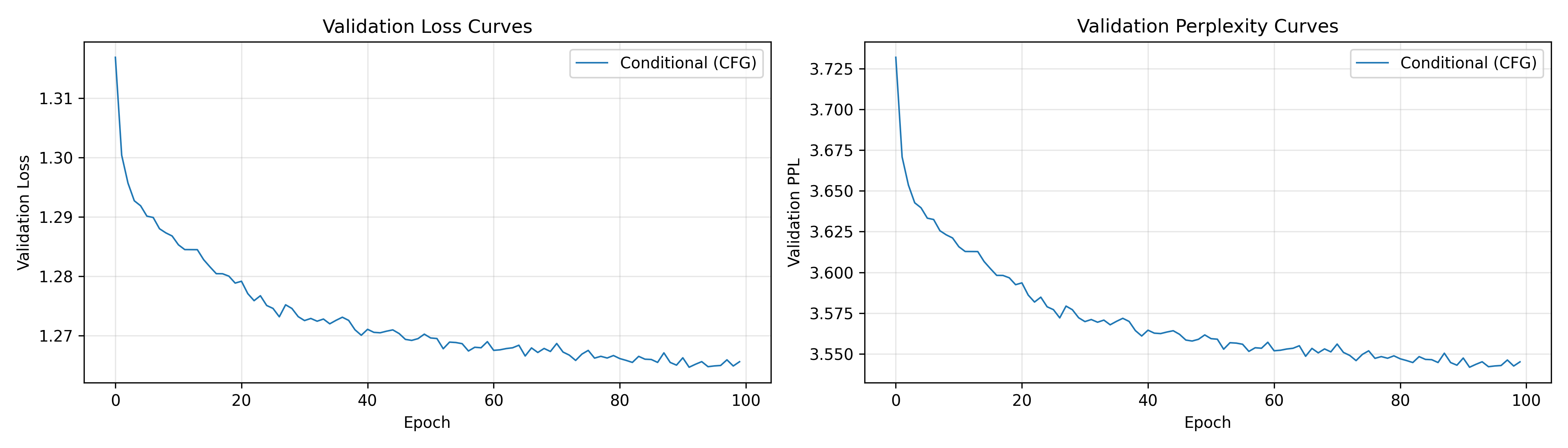} 
    \caption{Validation loss and perplexity of the class-conditioned DNA-CRAFT diffusion model over 100 training epochs.} 
    \label{fig:validation_curves} 
\end{figure}

\textbf{Computational Infrastructure.} All experiments were conducted on NVIDIA H100 GPUs (80GB VRAM). The validation loss and perplexity trajectories of the class-conditioned diffusion model are shown in Figure~\ref{fig:validation_curves}.

\subsection{Class-Conditioned Sampling and Motif Analysis}
\label{app:class_cond_details}

\textbf{Classifier-Free Guidance Renormalization.}
\label{app:cfg_renorm}
Equation~\ref{eq:cfg} applies guidance to the denoiser's clean-token probabilities, with normalization over nucleotides at each masked position. We then construct the reverse transition using the noise schedule $\alpha_t$ defined in Appendix~\ref{app:diffusion_training}. For a masked position $\ell$ and $s<t$, the probability of token $a$ is
\begin{equation}
\label{eq:CFG_parallel_sampling}
p_\theta^\gamma(\bm{z}_s^\ell=a\mid\bm{z}_t,\bm{y}) =
\begin{cases}
\displaystyle\frac{\alpha_s-\alpha_t}{1-\alpha_t}[\bm{x}_\theta^\gamma(\bm{z}_t,t,\bm{y})]_{\ell,a}, & a\in\mathcal{V},\\[5pt]
\displaystyle\frac{1-\alpha_s}{1-\alpha_t}, & a=\bm{m}.
\end{cases}
\end{equation}
Thus, guidance changes the nucleotide probabilities while the probability of remaining masked follows the diffusion schedule. At already unmasked positions, the transition preserves the current nucleotide. Conditional on the current sequence, sampling factorizes across positions and uses independent Gumbel draws as specified in Equation~\ref{eq:child_sampling}.

\begin{figure}[ht]
    \centering
    \includegraphics[width=0.7\linewidth]{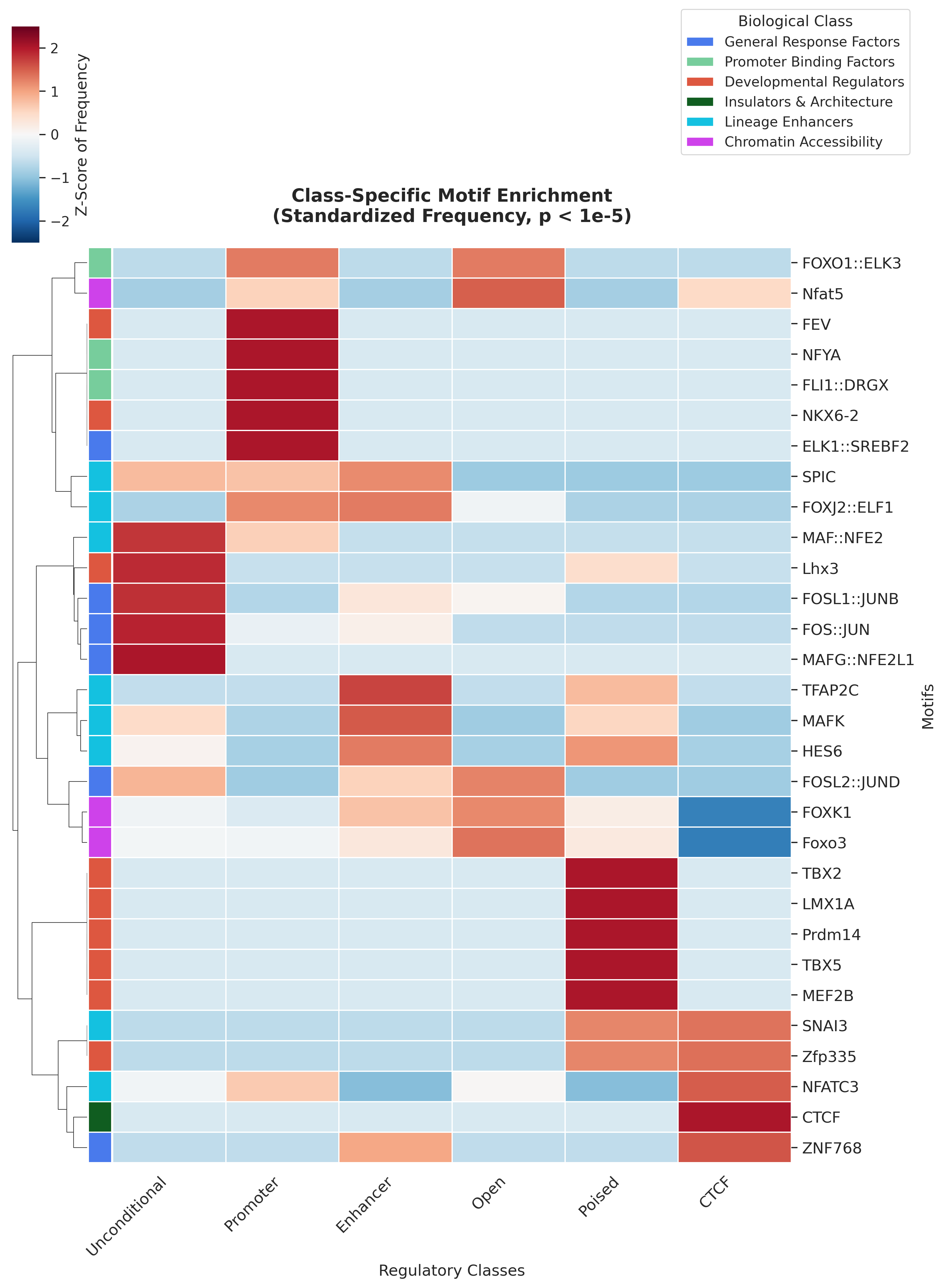}
    \caption{Heatmap of motif-frequency Z-scores across the five regulatory classes and null-conditioned sampling. ``Unconditional'' denotes sampling from the same class-conditioned network with the null label at $\gamma=0$.}
    \label{fig:detailed_motif_heatmap}
    \vspace{-15pt}
\end{figure}

\textbf{Detailed Motif Enrichment Analysis.} To validate the biological relevance of the class-conditioned generated sequences, we performed a motif enrichment analysis. Figure \ref{fig:detailed_motif_heatmap} displays a heatmap of the Z-scores for the top 5 most specific motifs per class. The recovery of these specific factors is consistent with known biological expectations. Table \ref{tab:top_motifs_class} lists the top 5 enriched TFs for each class.

\begin{table}[ht]
    \centering
    \caption{Top 5 Enriched Transcription Factors per Regulatory Class. Factors are ranked by enrichment Z-score.}
    \label{tab:top_motifs_class}
    \begin{tabular}{@{}lp{0.6\textwidth}@{}}
        \toprule
        \textbf{Regulatory Class} & \textbf{Top Enriched Motifs} \\
        \midrule
        \textbf{CTCF} & CTCF, ZNF768, NFATC3, Zfp335, SNAI3 \\
        \addlinespace[0.3em]
        \textbf{Enhancer} & TFAP2C, MAFK, HES6, FOXJ2::ELF1, SPIC \\
        \addlinespace[0.3em]
        \textbf{Open Chromatin} & Nfat5, Foxo3, FOXO1::ELK3, FOSL2::JUND, FOXK1 \\
        \addlinespace[0.3em]
        \textbf{Poised} & LMX1A, TBX5, Prdm14, MEF2B, TBX2 \\
        \addlinespace[0.3em]
        \textbf{Promoter} & NKX6-2, FEV, NFYA, FLI1::DRGX, ELK1::SREBF2 \\
        \addlinespace[0.3em]
        \textbf{Unconditional} & MAFG::NFE2L1, FOS::JUN, Lhx3, FOSL1::JUNB, MAF::NFE2 \\
        \bottomrule
    \end{tabular}
\end{table}

\subsection{Regulatory Sequence Design Benchmarks}
\label{app:benchmark_details}

\textbf{Benchmarking Model Implementation Details.} Reported benchmark values summarize three independent generation runs, each producing 128 sequences. Each run contributes one aggregate value per metric. Standard deviations describe variation across generation runs, and bold values indicate the highest or lowest reported means according to the metric direction.

\begin{enumerate}
    \item \textbf{Ledidi:} We initialized the optimization with 128 random DNA sequences. We implemented a custom differentiable wrapper to compute MinGap scores from Design-model's output and backpropagate gradients directly to the input sequence representation. Optimization was conducted independently for each sequence for a maximum of 20,000 steps to ensure convergence.
    \item \textbf{Sequential Monte Carlo (SMC):} We used the same class-conditioned diffusion model as DNA-CRAFT, conditioned on the Enhancer class with CFG scale $\gamma=1.0$. The sampling process tracked 128 particles and applied a resampling parameter of $\alpha=0.5$.
    \item \textbf{Twisted Diffusion Sampling (TDS):} We used the same enhancer-conditioned diffusion prior and particle configuration as SMC. TDS used a reward-guidance scale of 1000 and a resampling parameter of $\alpha=0.5$.
    \item \textbf{D3 (Discrete Denoising Diffusion):} We trained the discrete diffusion model utilizing the transformer backbone (2M parameters) for 100 epochs on the MPRA dataset (${\sim}$700,000 sequences) with the cell type activity as classes. We generated 128 sequences per cell type with conditional sampling ($\gamma=4.0$).
    \item \textbf{Ctrl-DNA:} We followed the original protocol and hyperparameters specified for each target cell type. We fine-tuned three separate models (one for each cell line: HepG2, K562, SK-N-SH) for 100 optimization steps. The top 128 sequences from each fine-tuning run were selected for evaluation. We note that Ctrl-DNA's base autoregressive model \citep{nguyen2023hyenadna} is trained on the entire human reference genome, which is a substantially larger and more diverse training corpus than the ENCODE registry of 3.2 million cCREs used to train DNA-CRAFT's diffusion backbone. Additionally, the TFBS regularization term was excluded, as the specific motif lists were not available in the public repository.
    \item \textbf{DRAKES:} We followed the protocol described in the original publication. For the HepG2 cell line, we utilized the provided pre-trained checkpoint. Similarly, we fine-tuned two separate models for K562 and SK-N-SH. We generated sequences using the respective fine-tuned models. We note that DRAKES fine-tunes its diffusion model using reward gradients computed over the full MPRA dataset (${\sim}$700,000 sequences), which encompasses the data used to train both our Design-model and Evaluation-model. This gives DRAKES implicit access to the evaluation distribution during its fine-tuning process.
    \item \textbf{DNA-CRAFT (Ours):} We use the class-conditioned diffusion model trained on ENCODE without further fine-tuning. For each replicate of the human cell-line benchmark, we initialize 128 separate search trees, each with a fresh archive and a fully masked root. Each search scores candidate completions using the Design model and maintains an archive according to the smooth specificity score $g$. We retain the highest-scoring completion from each search to obtain the final set of 128 sequences per replicate. Table~\ref{tab:dnacraft_params} summarizes the inference settings.
\end{enumerate}
\begin{table}[ht]
\centering
\caption{DNA-CRAFT inference parameters for the human cell-line benchmark. Each experimental replicate comprises 128 separately initialized searches.}
\label{tab:dnacraft_params}
\begin{tabular}{@{}ll@{}}
\toprule
\textbf{Parameter} & \textbf{Value} \\
\midrule
Diffusion Sampling Steps & 32 \\
Guidance Scale ($\gamma$) & 1.0 \\
Conditioning Class & Enhancer \\
Search Iterations ($N_{\text{iter}}$) & 32 \\
Retained-Child Limit ($M$) & 8 \\
Lookahead Pool Size ($B$) & 800 \\
UCT Exploration Coefficient ($c$) & 0.25 \\
Archive Capacity ($N_{\text{max}}$) & 64\\
Separate Searches per Replicate & 128 \\
\bottomrule
\end{tabular}
\end{table}

\textbf{Extended Benchmark Results}
\label{app:extended_benchmarks}

We extended our evaluation to include the following additional evaluation metrics:

\begin{itemize}
    \item \textbf{Complete Dataset Validation:} We additionally scored the generated sequences using an Enformer model trained on the full MPRA dataset \citep{gosai2024machine}, providing a supplementary comparison with the partition-trained Evaluation model.
    
    \item \textbf{Fr\'{e}chet Biological Distance (FBD):} Following \citet{stark2024dirichletflowmatchingapplications}, we use the Fr\'{e}chet Biological Distance (FBD) to measure the resemblance of generated sequences to the natural distribution. This metric extends the Fréchet Inception Distance (FID) commonly used in image generation~\citep{heusel2018ganstrainedtimescaleupdate} to biological sequences by computing distances in a learned embedding space. Concretely, let $\{\bm{h}_i^{\mathrm{ref}}\}_{i=1}^{N}$ and $\{\bm{h}_j^{\mathrm{gen}}\}_{j=1}^{M}$ denote embeddings of reference and generated sequences, respectively, obtained from the sequence-to-activity model. We fit Gaussian distributions $\mathcal{N}(\bm{\mu}_{\mathrm{ref}}, \bm{\Sigma}_{\mathrm{ref}})$ and $\mathcal{N}(\bm{\mu}_{\mathrm{gen}}, \bm{\Sigma}_{\mathrm{gen}})$ to these embeddings, and compute the Fréchet distance:
    \begin{equation}
    \mathrm{FBD} = 
    \|\bm{\mu}_{\mathrm{ref}} - \bm{\mu}_{\mathrm{gen}}\|_2^2
    + \mathrm{Tr}\!\left(
    \bm{\Sigma}_{\mathrm{ref}} + \bm{\Sigma}_{\mathrm{gen}}
    - 2\left(\bm{\Sigma}_{\mathrm{ref}}\bm{\Sigma}_{\mathrm{gen}}\right)^{1/2}
    \right).
    \end{equation}
    Lower FBD indicates closer alignment between the generated and reference sequence distributions. 
    \item \textbf{Pairwise Levenshtein Distance:} We calculated sequence diversity using pairwise Levenshtein (edit) distance.
    
    \item \textbf{Cross-Architecture Validation (MinGap Borzoi):} We evaluate the predicted activity of the generated sequences with a fine-tuned Borzoi model \citep{linder_predicting_2025} trained on the evaluation data split. This tests whether the cell-type specificity transfers across independent predictor architectures.

    \item \textbf{GC Content:} We report the mean GC fraction of each method's generated sequences. The reference value (shown in parentheses) is the mean GC content of natural enhancers for the target cell type. Methods that produce sequences with GC content far from the reference value may be exploiting compositional biases rather than learning genuine regulatory syntax.

    \item \textbf{Motif-Hit Preprocessing:} For the arrangement metrics, we restrict the analysis to the 50 most frequent motifs in the natural reference set and retain FIMO \citep{grant_fimo_2011} hits with $p < 10^{-4}$. Within each sequence, hits for the same motif are selected greedily by score, requiring at least 10\,bp between their centers. We then retain at most the 10 highest-scoring hits per sequence in both the generated and reference sets. The same reference-defined motif vocabulary is used for all methods.

    \item \textbf{Pairwise TFBS Co-occurrence:} For each unordered pair of distinct motifs, we compute the association $\log_2[(P(a,b))/(P(a)P(b))]$, where $P(a,b)$ is the fraction of sequences containing both motifs, and $P(a)$ and $P(b)$ are their marginal presence frequencies. We report the Spearman correlation between the generated and reference association vectors across all motif pairs. Higher correlation indicates closer agreement with the motif co-occurrence patterns of natural sequences.
    
    \item \textbf{Inter-Motif Spacing (JSD):} We collect center-to-center distances between all pairs of distinct-motif hits within each sequence and pool these distances across the sequence set. Distances are grouped into 10\,bp bins spanning the sequence length, giving 20 bins for 200\,bp sequences. We report the base-2 Jensen--Shannon divergence (JSD) between the normalized generated and reference histograms. Lower values indicate closer agreement with the reference spacing distribution.
    
    \item \textbf{Relative Motif Orientation:} For each unordered pair of distinct motifs, we calculate the fraction of hit pairs within 50\,bp that occur on the same strand. We retain motif pairs with at least five observations in both the generated and reference sets and report the Spearman correlation between their same-strand fractions. The metric is reported only when at least five motif pairs satisfy this criterion and both vectors have nonzero variance. Higher correlation indicates closer agreement with the reference orientation preferences.

    \item \textbf{Fr\'{e}chet Biological Distance (NTv2):} We compute a second distributional distance using embeddings from the Nucleotide Transformer v2 (500M parameters) \citep{dalla-torre_nucleotide_2025}. This provides a measure of how close the generated distribution is to the natural enhancer distribution. Lower values are better.
    
\end{itemize}

\begin{figure}[t]
    \includegraphics[width=\linewidth]{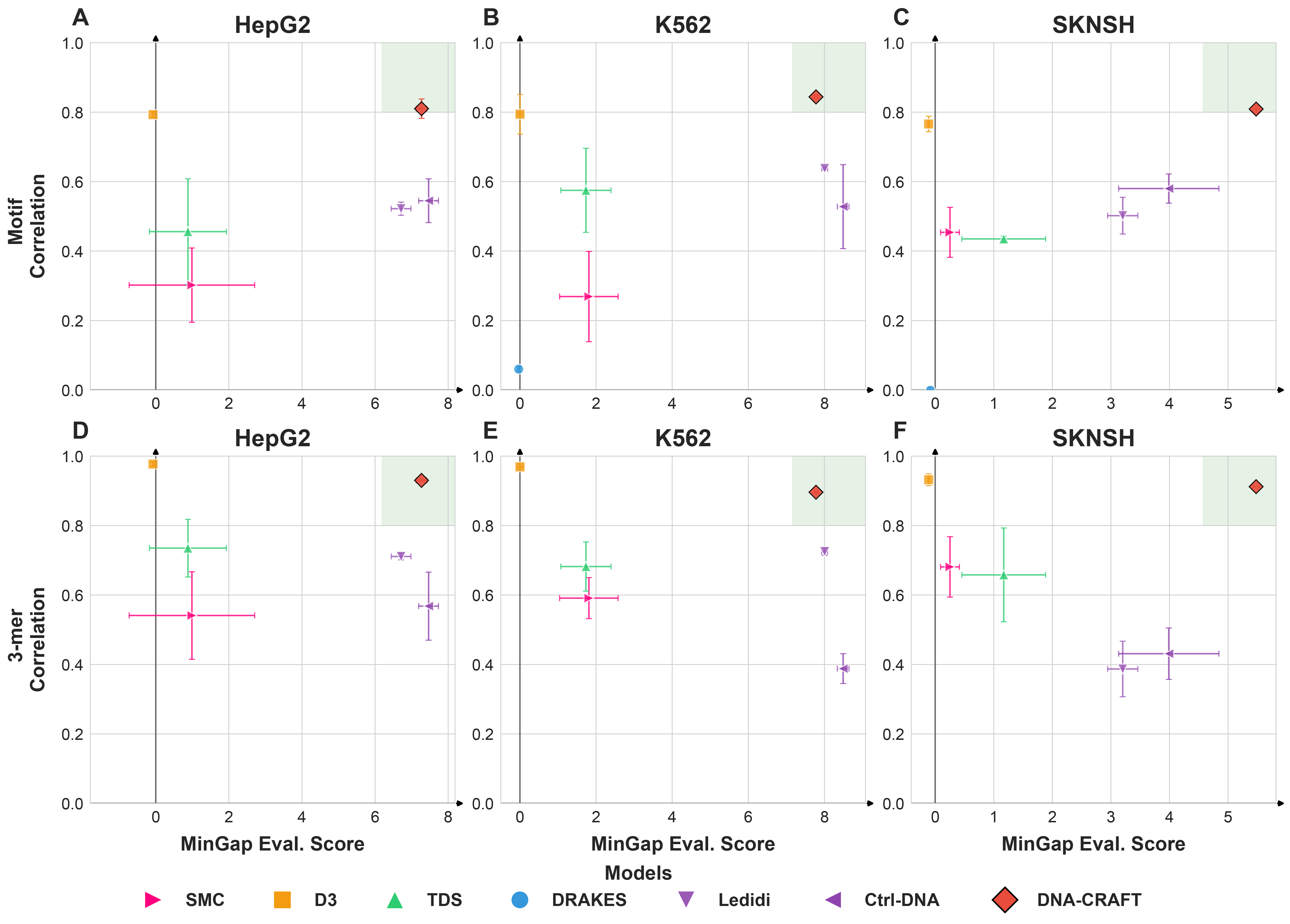}
    \caption{Trade-off between cell-type specificity and biological fidelity. 
    Performance of DNA-CRAFT compared to baselines for HepG2, K562, and SK-N-SH cell lines. The x-axis represents the MinGap score, serving as a proxy for cell-type-specific activity. The y-axis represents biological fidelity, measured by JASPAR Motif Correlation and 3-mer Correlation relative to top specific natural enhancers.}
    \label{fig:tradeoff_benchmarks}
\end{figure}

Table \ref{tab:human_enhancer_extended} presents the comprehensive performance across all metrics. Figure \ref{fig:tradeoff_benchmarks} visualizes the trade-off between optimization intensity (MinGap Eval. score) and naturalness. The figure illustrates the trade-offs between predicted specificity and the two fidelity metrics across methods and cell lines.

\begin{table}[t]
\caption{Extended human cell-line benchmarks. We report cell-type specificity under three oracle configurations spanning two architectures (Enformer evaluation-split, Enformer full-MPRA, and Borzoi), sequence-composition fidelity (motif correlation, 3-mer correlation, GC content), motif arrangement quality (pairwise co-occurrence, inter-motif spacing JSD, relative orientation correlation), distributional distance to natural enhancers (FBD with Enformer embeddings and FBD with Nucleotide Transformer v2 embeddings), and output diversity (pairwise Levenshtein distance and positional Shannon entropy). Arrows indicate the direction of improvement; GC content lists the natural-enhancer reference value. Best values are \textbf{bolded}. All values report mean $\pm$ std over three seeds.}
\label{tab:human_enhancer_extended}
\vspace{0.5em}
\centering
\resizebox{\textwidth}{!}{
\begin{tabular}{ll ccccccc}
\toprule
\textbf{Cell Type} & \textbf{Metric} & \textbf{SMC} & \textbf{TDS} & \textbf{D3} & \textbf{DRAKES} & \textbf{Ctrl-DNA} & \textbf{Ledidi} & \textbf{DNA-CRAFT (Ours)} \\
\midrule
\multirow{13}{*}{\textbf{HepG2}} & MinGap Eval $\uparrow$ & 0.989 \scriptsize{$\pm 1.717$} & 0.880 \scriptsize{$\pm 1.053$} & -0.075 \scriptsize{$\pm 0.021$} & -1.314 \scriptsize{$\pm 0.024$} & \textbf{7.464} \scriptsize{$\pm 0.270$} & 6.712 \scriptsize{$\pm 0.270$} & 7.271 \scriptsize{$\pm 0.142$} \\
 & MinGap (Full MPRA) $\uparrow$ & 0.875 \scriptsize{$\pm 1.575$} & 0.896 \scriptsize{$\pm 0.982$} & -0.056 \scriptsize{$\pm 0.014$} & -1.210 \scriptsize{$\pm 0.035$} & \textbf{7.210} \scriptsize{$\pm 0.714$} & 6.487 \scriptsize{$\pm 0.272$} & 7.127 \scriptsize{$\pm 0.168$} \\
 & MinGap (Borzoi) $\uparrow$ & 0.690 \scriptsize{$\pm 1.161$} & 0.601 \scriptsize{$\pm 0.791$} & -0.060 \scriptsize{$\pm 0.011$} & -1.100 \scriptsize{$\pm 0.040$} & \textbf{6.961} \scriptsize{$\pm 0.614$} & 4.946 \scriptsize{$\pm 0.159$} & 5.515 \scriptsize{$\pm 0.184$} \\
 & Motif Corr. $\uparrow$ & 0.302 \scriptsize{$\pm 0.107$} & 0.456 \scriptsize{$\pm 0.152$} & 0.793 \scriptsize{$\pm 0.009$} & -0.126 \scriptsize{$\pm 0.014$} & 0.545 \scriptsize{$\pm 0.063$} & 0.522 \scriptsize{$\pm 0.019$} & \textbf{0.810} \scriptsize{$\pm 0.028$} \\
 & 3-mer Corr. $\uparrow$ & 0.541 \scriptsize{$\pm 0.126$} & 0.735 \scriptsize{$\pm 0.083$} & \textbf{0.977} \scriptsize{$\pm 0.006$} & -0.361 \scriptsize{$\pm 0.012$} & 0.568 \scriptsize{$\pm 0.098$} & 0.711 \scriptsize{$\pm 0.010$} & 0.930 \scriptsize{$\pm 0.003$} \\
 & GC Content (ref. 0.457) & 0.485 \scriptsize{$\pm 0.060$} & 0.467 \scriptsize{$\pm 0.021$} & 0.458 \scriptsize{$\pm 0.009$} & 0.698 \scriptsize{$\pm 0.002$} & 0.458 \scriptsize{$\pm 0.032$} & 0.477 \scriptsize{$\pm 0.002$} & 0.487 \scriptsize{$\pm 0.001$} \\
 & Motif Co-occur. $\uparrow$ & 0.090 \scriptsize{$\pm 0.044$} & 0.145 \scriptsize{$\pm 0.039$} & 0.271 \scriptsize{$\pm 0.016$} & 0.225 \scriptsize{$\pm 0.046$} & 0.208 \scriptsize{$\pm 0.086$} & 0.304 \scriptsize{$\pm 0.024$} & \textbf{0.326} \scriptsize{$\pm 0.035$} \\
 & Spacing JSD $\downarrow$ & 0.123 \scriptsize{$\pm 0.025$} & 0.161 \scriptsize{$\pm 0.058$} & 0.022 \scriptsize{$\pm 0.006$} & 0.037 \scriptsize{$\pm 0.012$} & 0.022 \scriptsize{$\pm 0.002$} & \textbf{0.009} \scriptsize{$\pm 0.001$} & 0.011 \scriptsize{$\pm 0.001$} \\
 & Orientation Corr. $\uparrow$ & 0.269 \scriptsize{$\pm 0.023$} & 0.319 \scriptsize{$\pm 0.469$} & 0.452 \scriptsize{$\pm 0.147$} & 0.686 \scriptsize{$\pm 0.198$} & 0.455 \scriptsize{$\pm 0.115$} & \textbf{0.765} \scriptsize{$\pm 0.003$} & 0.695 \scriptsize{$\pm 0.026$} \\
 & FBD $\downarrow$ & 941.851 \scriptsize{$\pm 356.067$} & \textbf{904.089} \scriptsize{$\pm 172.563$} & 1066.242 \scriptsize{$\pm 14.740$} & 2816.206 \scriptsize{$\pm 101.942$} & 2968.966 \scriptsize{$\pm 776.318$} & 1535.551 \scriptsize{$\pm 70.126$} & 1427.015 \scriptsize{$\pm 127.205$} \\
 & FBD (NTv2) $\downarrow$ & 5.558 \scriptsize{$\pm 0.762$} & 4.210 \scriptsize{$\pm 0.450$} & \textbf{1.629} \scriptsize{$\pm 0.013$} & 13.231 \scriptsize{$\pm 1.080$} & 3.745 \scriptsize{$\pm 0.352$} & 2.397 \scriptsize{$\pm 0.034$} & 2.018 \scriptsize{$\pm 0.090$} \\
 & Diversity (Levenshtein) $\uparrow$ & 71.916 \scriptsize{$\pm 41.709$} & 104.328 \scriptsize{$\pm 8.925$} & 149.687 \scriptsize{$\pm 0.217$} & 141.970 \scriptsize{$\pm 0.118$} & 145.024 \scriptsize{$\pm 2.293$} & 149.776 \scriptsize{$\pm 0.145$} & \textbf{149.898} \scriptsize{$\pm 0.041$} \\
 & Shannon Div. $\uparrow$ & 0.845 \scriptsize{$\pm 0.478$} & 1.301 \scriptsize{$\pm 0.105$} & 1.978 \scriptsize{$\pm 0.003$} & 1.864 \scriptsize{$\pm 0.002$} & 1.914 \scriptsize{$\pm 0.033$} & 1.980 \scriptsize{$\pm 0.002$} & \textbf{1.981} \scriptsize{$\pm 0.001$} \\
\midrule
\multirow{13}{*}{\textbf{K562}} & MinGap Eval $\uparrow$ & 1.812 \scriptsize{$\pm 0.768$} & 1.735 \scriptsize{$\pm 0.660$} & 0.003 \scriptsize{$\pm 0.070$} & -0.033 \scriptsize{$\pm 0.032$} & \textbf{8.490} \scriptsize{$\pm 0.153$} & 8.004 \scriptsize{$\pm 0.078$} & 7.777 \scriptsize{$\pm 0.054$} \\
 & MinGap (Full MPRA) $\uparrow$ & 1.963 \scriptsize{$\pm 0.948$} & 1.752 \scriptsize{$\pm 0.714$} & -0.014 \scriptsize{$\pm 0.059$} & -0.275 \scriptsize{$\pm 0.071$} & \textbf{8.788} \scriptsize{$\pm 0.166$} & 8.324 \scriptsize{$\pm 0.099$} & 8.046 \scriptsize{$\pm 0.073$} \\
 & MinGap (Borzoi) $\uparrow$ & 1.123 \scriptsize{$\pm 0.845$} & 1.426 \scriptsize{$\pm 0.585$} & -0.042 \scriptsize{$\pm 0.049$} & -0.027 \scriptsize{$\pm 0.047$} & \textbf{7.656} \scriptsize{$\pm 0.050$} & 6.935 \scriptsize{$\pm 0.082$} & 6.747 \scriptsize{$\pm 0.095$} \\
 & Motif Corr. $\uparrow$ & 0.269 \scriptsize{$\pm 0.130$} & 0.575 \scriptsize{$\pm 0.121$} & 0.794 \scriptsize{$\pm 0.057$} & 0.060 \scriptsize{$\pm 0.008$} & 0.528 \scriptsize{$\pm 0.121$} & 0.638 \scriptsize{$\pm 0.004$} & \textbf{0.844} \scriptsize{$\pm 0.005$} \\
 & 3-mer Corr. $\uparrow$ & 0.591 \scriptsize{$\pm 0.059$} & 0.682 \scriptsize{$\pm 0.071$} & \textbf{0.969} \scriptsize{$\pm 0.013$} & -0.354 \scriptsize{$\pm 0.007$} & 0.388 \scriptsize{$\pm 0.043$} & 0.725 \scriptsize{$\pm 0.011$} & 0.896 \scriptsize{$\pm 0.003$} \\
 & GC Content (ref. 0.477) & 0.457 \scriptsize{$\pm 0.048$} & 0.499 \scriptsize{$\pm 0.067$} & 0.468 \scriptsize{$\pm 0.010$} & 0.567 \scriptsize{$\pm 0.003$} & 0.480 \scriptsize{$\pm 0.035$} & 0.488 \scriptsize{$\pm 0.000$} & 0.497 \scriptsize{$\pm 0.002$} \\
 & Motif Co-occur. $\uparrow$ & 0.105 \scriptsize{$\pm 0.065$} & 0.090 \scriptsize{$\pm 0.018$} & \textbf{0.409} \scriptsize{$\pm 0.024$} & 0.341 \scriptsize{$\pm 0.027$} & 0.069 \scriptsize{$\pm 0.041$} & 0.232 \scriptsize{$\pm 0.046$} & 0.383 \scriptsize{$\pm 0.059$} \\
 & Spacing JSD $\downarrow$ & 0.219 \scriptsize{$\pm 0.044$} & 0.184 \scriptsize{$\pm 0.078$} & 0.021 \scriptsize{$\pm 0.003$} & 0.025 \scriptsize{$\pm 0.008$} & 0.053 \scriptsize{$\pm 0.016$} & 0.013 \scriptsize{$\pm 0.003$} & \textbf{0.012} \scriptsize{$\pm 0.001$} \\
 & Orientation Corr. $\uparrow$ & 0.674 \scriptsize{$\pm 0.183$} & 0.866 \scriptsize{$\pm 0.038$} & 0.713 \scriptsize{$\pm 0.171$} & 0.664 \scriptsize{$\pm 0.058$} & 0.759 \scriptsize{$\pm 0.074$} & 0.764 \scriptsize{$\pm 0.080$} & \textbf{0.869} \scriptsize{$\pm 0.033$} \\
 & FBD $\downarrow$ & \textbf{742.185} \scriptsize{$\pm 94.080$} & 799.443 \scriptsize{$\pm 176.784$} & 1145.004 \scriptsize{$\pm 31.744$} & 1962.765 \scriptsize{$\pm 69.462$} & 2890.187 \scriptsize{$\pm 364.824$} & 1853.457 \scriptsize{$\pm 33.499$} & 1619.139 \scriptsize{$\pm 18.759$} \\
 & FBD (NTv2) $\downarrow$ & 6.072 \scriptsize{$\pm 0.943$} & 4.724 \scriptsize{$\pm 0.313$} & \textbf{1.667} \scriptsize{$\pm 0.023$} & 8.666 \scriptsize{$\pm 0.638$} & 4.911 \scriptsize{$\pm 0.356$} & 2.264 \scriptsize{$\pm 0.062$} & 2.031 \scriptsize{$\pm 0.062$} \\
 & Diversity (Levenshtein) $\uparrow$ & 39.905 \scriptsize{$\pm 27.443$} & 92.245 \scriptsize{$\pm 10.923$} & 149.809 \scriptsize{$\pm 0.171$} & 148.352 \scriptsize{$\pm 0.213$} & 141.204 \scriptsize{$\pm 0.978$} & \textbf{149.896} \scriptsize{$\pm 0.084$} & 149.789 \scriptsize{$\pm 0.051$} \\
 & Shannon Div. $\uparrow$ & 0.487 \scriptsize{$\pm 0.302$} & 1.063 \scriptsize{$\pm 0.170$} & 1.980 \scriptsize{$\pm 0.002$} & 1.958 \scriptsize{$\pm 0.003$} & 1.851 \scriptsize{$\pm 0.019$} & \textbf{1.982} \scriptsize{$\pm 0.001$} & 1.980 \scriptsize{$\pm 0.001$} \\
\midrule
\multirow{13}{*}{\textbf{SK-N-SH}} & MinGap Eval $\uparrow$ & 0.252 \scriptsize{$\pm 0.161$} & 1.170 \scriptsize{$\pm 0.715$} & -0.111 \scriptsize{$\pm 0.014$} & -0.082 \scriptsize{$\pm 0.040$}& 3.988 \scriptsize{$\pm 0.856$} & 3.202 \scriptsize{$\pm 0.258$} & \textbf{5.479} \scriptsize{$\pm 0.057$} \\
 & MinGap (Full MPRA) $\uparrow$ & 0.222 \scriptsize{$\pm 0.133$} & 1.125 \scriptsize{$\pm 0.779$} & -0.116 \scriptsize{$\pm 0.011$} & 0.255 \scriptsize{$\pm 0.051$} & 4.598 \scriptsize{$\pm 0.843$} & 3.680 \scriptsize{$\pm 0.370$} & \textbf{6.894} \scriptsize{$\pm 0.102$} \\
 & MinGap (Borzoi) $\uparrow$ & 0.070 \scriptsize{$\pm 0.191$} & 0.723 \scriptsize{$\pm 0.580$} & -0.158 \scriptsize{$\pm 0.040$} & -0.036 \scriptsize{$\pm 0.040$} & 3.650 \scriptsize{$\pm 0.886$} & 2.431 \scriptsize{$\pm 0.203$} & \textbf{4.565} \scriptsize{$\pm 0.071$} \\
 & Motif Corr. $\uparrow$ & 0.454 \scriptsize{$\pm 0.072$} & 0.435 \scriptsize{$\pm 0.008$} & 0.766 \scriptsize{$\pm 0.022$} & -0.001 \scriptsize{$\pm 0.009$} & 0.580 \scriptsize{$\pm 0.042$} & 0.502 \scriptsize{$\pm 0.053$} & \textbf{0.809} \scriptsize{$\pm 0.010$} \\
 & 3-mer Corr. $\uparrow$ & 0.681 \scriptsize{$\pm 0.087$} & 0.658 \scriptsize{$\pm 0.135$} & \textbf{0.932} \scriptsize{$\pm 0.017$} & -0.382 \scriptsize{$\pm 0.001$} & 0.431 \scriptsize{$\pm 0.074$} & 0.387 \scriptsize{$\pm 0.080$} & 0.912 \scriptsize{$\pm 0.001$} \\
 & GC Content (ref. 0.462) & 0.420 \scriptsize{$\pm 0.059$} & 0.475 \scriptsize{$\pm 0.080$} & 0.433 \scriptsize{$\pm 0.005$} & 0.720 \scriptsize{$\pm 0.001$} & 0.477 \scriptsize{$\pm 0.063$} & 0.487 \scriptsize{$\pm 0.005$} & 0.440 \scriptsize{$\pm 0.001$} \\
 & Motif Co-occur. $\uparrow$ & 0.039 \scriptsize{$\pm 0.176$} & 0.065 \scriptsize{$\pm 0.121$} & 0.432 \scriptsize{$\pm 0.038$} & 0.140 \scriptsize{$\pm 0.122$} & 0.428 \scriptsize{$\pm 0.080$} & 0.481 \scriptsize{$\pm 0.016$} & \textbf{0.534} \scriptsize{$\pm 0.036$} \\
 & Spacing JSD $\downarrow$ & 0.170 \scriptsize{$\pm 0.092$} & 0.222 \scriptsize{$\pm 0.096$} & 0.033 \scriptsize{$\pm 0.009$} & 0.066 \scriptsize{$\pm 0.031$} & 0.042 \scriptsize{$\pm 0.050$} & 0.021 \scriptsize{$\pm 0.011$} & \textbf{0.008} \scriptsize{$\pm 0.002$} \\
 & Orientation Corr. $\uparrow$ & 0.279 \scriptsize{$\pm 0.720$} & 0.336 \scriptsize{$\pm 0.321$} & 0.711 \scriptsize{$\pm 0.098$} & $-$ & 0.871 \scriptsize{$\pm 0.054$} & 0.891 \scriptsize{$\pm 0.038$} & \textbf{0.897} \scriptsize{$\pm 0.016$} \\
 & FBD $\downarrow$ & 1047.505 \scriptsize{$\pm 23.097$} & 841.456 \scriptsize{$\pm 85.692$} & 1016.487 \scriptsize{$\pm 15.447$} & 2291.623 \scriptsize{$\pm 35.281$} & 1834.831 \scriptsize{$\pm 1284.741$} & \textbf{777.882} \scriptsize{$\pm 75.112$} & 1742.182 \scriptsize{$\pm 44.163$} \\
 & FBD (NTv2) $\downarrow$ & 5.893 \scriptsize{$\pm 1.737$} & 5.538 \scriptsize{$\pm 0.686$} & \textbf{1.785} \scriptsize{$\pm 0.016$} & 18.779 \scriptsize{$\pm 0.856$} & 4.866 \scriptsize{$\pm 0.750$} & 2.687 \scriptsize{$\pm 0.104$} & 2.357 \scriptsize{$\pm 0.013$} \\
 & Diversity (Levenshtein) $\uparrow$ & 81.457 \scriptsize{$\pm 8.492$} & 82.683 \scriptsize{$\pm 28.917$} & 149.011 \scriptsize{$\pm 0.246$} & 139.740 \scriptsize{$\pm 0.052$} & 140.473 \scriptsize{$\pm 4.902$} & \textbf{149.944} \scriptsize{$\pm 0.067$} & 149.218 \scriptsize{$\pm 0.069$} \\
 & Shannon Div. $\uparrow$ & 1.029 \scriptsize{$\pm 0.091$} & 1.019 \scriptsize{$\pm 0.327$} & 1.969 \scriptsize{$\pm 0.004$} & 1.826 \scriptsize{$\pm 0.001$} & 1.847 \scriptsize{$\pm 0.066$} & \textbf{1.982} \scriptsize{$\pm 0.001$} & 1.972 \scriptsize{$\pm 0.001$} \\
\bottomrule
\end{tabular}
}
\end{table}

\subsubsection{Brain Cell-Type-Specific Sequence Design}
\label{app:brain_benchmark}

\textbf{Experimental Setup.} We further evaluated regulatory sequence design across eight brain cell types from the CATlas dataset: Astro, GABA1, GABA2, Glut1, Glut2, Microglia, OPC, and Oligo. We followed the split-model protocol described in Section~\ref{sec:immune_benchmark}, using separate Design and Evaluation models trained on non-overlapping homology clusters. Each experiment targeted one cell type while penalizing predicted accessibility in the remaining seven. We compared SMC, TDS, Ctrl-DNA, Ledidi, and DNA-CRAFT, generating 128 sequences per run over three independent random seeds. DNA-CRAFT, SMC, and TDS used the same enhancer-conditioned diffusion prior with class-conditioning scale $\gamma=1.0$, and DNA-CRAFT used a lookahead pool of $B=800$. Evaluation followed the immune benchmark, reporting MinGap under the Evaluation model, motif correlation, 3-mer correlation, and Shannon diversity. Natural reference regions were selected using the same procedure as for the immune cell types.

\textbf{Results.} DNA-CRAFT achieves the highest reported mean MinGap, motif correlation, and 3-mer correlation across all eight brain cell types (Table~\ref{tab:brain_enhancer_results}). Its MinGap ranges from $0.408$ for Glut1 to $0.949$ for Microglia, while maintaining strong agreement with the reference motif and sequence-composition distributions. Ledidi achieves the highest positional Shannon diversity across all eight cell types, but substantially lower specificity and motif correlation. These results extend the observed specificity-naturalness trade-off to neuronal and glial cell types within the CATlas benchmark.

\begin{table}[h]
\centering
\caption{Comparison of methods for cell-type-specific regulatory sequence design across eight brain cell types. Shown are MinGap specificity scores from the Evaluation model, motif Spearman correlations, 3-mer Pearson correlations, and Shannon diversity. Values are reported as mean $\pm$ standard deviation over three independent runs, with 128 sequences per run. The highest reported mean in each row is marked in \textbf{bold}. Ctrl-DNA was evaluated without TFBS regularization.}
\label{tab:brain_enhancer_results}
\footnotesize
\setlength{\tabcolsep}{1pt}
\begin{tabular}{@{}llccccc@{}}
\toprule
\shortstack{\textbf{Target Cell}\\\textbf{Type}} & \textbf{Metric} & \textbf{SMC} & \textbf{TDS} & \textbf{Ctrl-DNA} & \textbf{Ledidi} & \shortstack{\textbf{DNA-CRAFT}\\\textbf{(Ours)}} \\
\midrule
\multirow{4}{*}{\textbf{Astro}} & MinGap $\uparrow$ & $-0.094 \pm 0.079$ & $-0.045 \pm 0.033$ & $0.311 \pm 0.164$ & $-0.180 \pm 0.004$ & $\mathbf{0.785} \pm 0.003$ \\
 & Motif Corr. $\uparrow$ & $0.389 \pm 0.086$ & $0.418 \pm 0.153$ & $0.372 \pm 0.209$ & $-0.057 \pm 0.119$ & $\mathbf{0.750} \pm 0.022$ \\
 & 3-mer Corr. $\uparrow$ & $0.699 \pm 0.110$ & $0.747 \pm 0.078$ & $0.498 \pm 0.117$ & $0.029 \pm 0.161$ & $\mathbf{0.914} \pm 0.002$ \\
 & Shannon Div. $\uparrow$ & $1.131 \pm 0.130$ & $1.263 \pm 0.172$ & $1.811 \pm 0.137$ & $\mathbf{1.983} \pm 0.001$ & $1.982 \pm 0.001$ \\
\midrule
\multirow{4}{*}{\textbf{GABA1}} & MinGap $\uparrow$ & $-0.069 \pm 0.006$ & $-0.035 \pm 0.030$ & $0.354 \pm 0.059$ & $-0.064 \pm 0.002$ & $\mathbf{0.638} \pm 0.010$ \\
 & Motif Corr. $\uparrow$ & $0.383 \pm 0.206$ & $0.331 \pm 0.257$ & $0.724 \pm 0.012$ & $-0.050 \pm 0.147$ & $\mathbf{0.839} \pm 0.002$ \\
 & 3-mer Corr. $\uparrow$ & $0.738 \pm 0.050$ & $0.729 \pm 0.099$ & $0.601 \pm 0.027$ & $0.009 \pm 0.215$ & $\mathbf{0.869} \pm 0.002$ \\
 & Shannon Div. $\uparrow$ & $1.297 \pm 0.116$ & $1.455 \pm 0.048$ & $1.929 \pm 0.011$ & $\mathbf{1.983} \pm 0.001$ & $1.965 \pm 0.001$ \\
\midrule
\multirow{4}{*}{\textbf{GABA2}} & MinGap $\uparrow$ & $-0.078 \pm 0.005$ & $-0.080 \pm 0.051$ & $0.592 \pm 0.039$ & $-0.072 \pm 0.001$ & $\mathbf{0.793} \pm 0.012$ \\
 & Motif Corr. $\uparrow$ & $0.510 \pm 0.179$ & $0.396 \pm 0.076$ & $0.717 \pm 0.034$ & $0.002 \pm 0.054$ & $\mathbf{0.737} \pm 0.017$ \\
 & 3-mer Corr. $\uparrow$ & $0.772 \pm 0.079$ & $0.692 \pm 0.068$ & $0.614 \pm 0.086$ & $0.020 \pm 0.197$ & $\mathbf{0.872} \pm 0.006$ \\
 & Shannon Div. $\uparrow$ & $1.510 \pm 0.063$ & $1.047 \pm 0.481$ & $1.908 \pm 0.014$ & $\mathbf{1.983} \pm 0.001$ & $1.946 \pm 0.001$ \\
\midrule
\multirow{4}{*}{\textbf{Glut1}} & MinGap $\uparrow$ & $-0.001 \pm 0.022$ & $0.022 \pm 0.025$ & $0.004 \pm 0.001$ & $0.034 \pm 0.003$ & $\mathbf{0.408} \pm 0.004$ \\
 & Motif Corr. $\uparrow$ & $0.541 \pm 0.092$ & $0.496 \pm 0.019$ & $0.588 \pm 0.050$ & $0.058 \pm 0.137$ & $\mathbf{0.731} \pm 0.017$ \\
 & 3-mer Corr. $\uparrow$ & $0.822 \pm 0.025$ & $0.781 \pm 0.068$ & $0.422 \pm 0.082$ & $0.054 \pm 0.169$ & $\mathbf{0.831} \pm 0.004$ \\
 & Shannon Div. $\uparrow$ & $1.476 \pm 0.121$ & $1.370 \pm 0.221$ & $1.844 \pm 0.033$ & $\mathbf{1.983} \pm 0.001$ & $1.976 \pm 0.001$ \\
\midrule
\multirow{4}{*}{\textbf{Glut2}} & MinGap $\uparrow$ & $-0.027 \pm 0.010$ & $-0.075 \pm 0.007$ & $0.028 \pm 0.003$ & $-0.064 \pm 0.001$ & $\mathbf{0.495} \pm 0.010$ \\
 & Motif Corr. $\uparrow$ & $0.401 \pm 0.053$ & $0.496 \pm 0.066$ & $0.687 \pm 0.042$ & $0.017 \pm 0.087$ & $\mathbf{0.760} \pm 0.018$ \\
 & 3-mer Corr. $\uparrow$ & $0.755 \pm 0.060$ & $0.760 \pm 0.098$ & $0.395 \pm 0.041$ & $0.011 \pm 0.215$ & $\mathbf{0.771} \pm 0.000$ \\
 & Shannon Div. $\uparrow$ & $1.327 \pm 0.147$ & $1.303 \pm 0.036$ & $1.737 \pm 0.045$ & $\mathbf{1.983} \pm 0.001$ & $1.895 \pm 0.004$ \\
\midrule
\multirow{4}{*}{\textbf{Microglia}} & MinGap $\uparrow$ & $-0.081 \pm 0.068$ & $-0.098 \pm 0.022$ & $0.786 \pm 0.093$ & $-0.095 \pm 0.011$ & $\mathbf{0.949} \pm 0.002$ \\
 & Motif Corr. $\uparrow$ & $0.241 \pm 0.149$ & $0.447 \pm 0.148$ & $0.706 \pm 0.007$ & $0.144 \pm 0.200$ & $\mathbf{0.858} \pm 0.004$ \\
 & 3-mer Corr. $\uparrow$ & $0.682 \pm 0.068$ & $0.688 \pm 0.017$ & $0.517 \pm 0.089$ & $0.116 \pm 0.145$ & $\mathbf{0.900} \pm 0.010$ \\
 & Shannon Div. $\uparrow$ & $1.203 \pm 0.366$ & $1.312 \pm 0.341$ & $1.850 \pm 0.049$ & $\mathbf{1.983} \pm 0.001$ & $1.973 \pm 0.002$ \\
\midrule
\multirow{4}{*}{\textbf{OPC}} & MinGap $\uparrow$ & $-0.072 \pm 0.025$ & $-0.058 \pm 0.018$ & $0.088 \pm 0.124$ & $-0.053 \pm 0.002$ & $\mathbf{0.577} \pm 0.003$ \\
 & Motif Corr. $\uparrow$ & $0.490 \pm 0.041$ & $0.418 \pm 0.170$ & $0.643 \pm 0.076$ & $0.017 \pm 0.116$ & $\mathbf{0.781} \pm 0.011$ \\
 & 3-mer Corr. $\uparrow$ & $0.761 \pm 0.103$ & $0.754 \pm 0.036$ & $0.535 \pm 0.092$ & $0.049 \pm 0.128$ & $\mathbf{0.904} \pm 0.003$ \\
 & Shannon Div. $\uparrow$ & $1.292 \pm 0.055$ & $1.328 \pm 0.032$ & $1.665 \pm 0.084$ & $\mathbf{1.983} \pm 0.001$ & $1.971 \pm 0.002$ \\
\midrule
\multirow{4}{*}{\textbf{Oligo}} & MinGap $\uparrow$ & $-0.103 \pm 0.061$ & $-0.095 \pm 0.026$ & $0.554 \pm 0.489$ & $-0.065 \pm 0.011$ & $\mathbf{0.907} \pm 0.003$ \\
 & Motif Corr. $\uparrow$ & $0.271 \pm 0.117$ & $0.389 \pm 0.108$ & $0.291 \pm 0.105$ & $0.039 \pm 0.134$ & $\mathbf{0.806} \pm 0.009$ \\
 & 3-mer Corr. $\uparrow$ & $0.620 \pm 0.136$ & $0.754 \pm 0.110$ & $0.329 \pm 0.097$ & $0.064 \pm 0.172$ & $\mathbf{0.884} \pm 0.013$ \\
 & Shannon Div. $\uparrow$ & $1.315 \pm 0.254$ & $1.384 \pm 0.146$ & $1.836 \pm 0.020$ & $\mathbf{1.983} \pm 0.001$ & $1.976 \pm 0.003$ \\
\bottomrule
\end{tabular}
\end{table}

\subsubsection{Sequence-To-Activity Model Training Details.} We adapted the pre-trained Enformer \citep{enformer} and Borzoi \citep{linder_predicting_2025} architectures using gReLU \citep{lal_grelu_2025}. For each task, we replaced the output head and fine-tuned the full network. Training used the Adam optimizer with a learning rate of $1 \times 10^{-4}$ and a batch size of 1024 for 15 epochs, without reverse-complement or shift augmentation. For the MPRA benchmark, the models predict activity across HepG2, K562, and SK-N-SH using a three-output regression head optimized with mean squared error. For chromatin accessibility prediction, the models use a 204-output head trained with binary cross-entropy loss.

\textbf{Input Processing and Prediction Scale.} Enformer receives the 200\,bp sequence directly. For Borzoi, we pad each sequence with 28 ambiguous bases on either side to obtain a 256\,bp input, using the same processing during training and evaluation. These inputs are shorter than the contexts used by the original pre-trained architectures. MPRA predictions are used directly on the trained regression scale. For chromatin accessibility, we apply a sigmoid to the output logits before calculating specificity, so the immune MinGap scores represent differences in predicted accessibility probabilities.

\textbf{Validation Performance.} For the partitioned Design and Evaluation models, training and validation use opposite data partitions. The Design model is trained on the design partition and validated on the evaluation partition, while the Evaluation model uses the reverse assignment. We select the checkpoint with the best validation performance for final scoring. Consequently, the reported scores summarize checkpoint-selection validation performance rather than performance on an additional untouched test set. Table~\ref{tab:oracle_performance} reports the MPRA validation metrics for Enformer and Borzoi, alongside the original Enformer Full-Model results. Table~\ref{tab:oracle_immune_performance} reports accessibility-prediction AUROC for the eight immune cell types used in the design benchmark and the mean across all 204 tracks.

\begin{table}[h]
\centering
\caption{Validation performance of sequence-to-activity models on the MPRA benchmark. Design and Evaluation models are evaluated on the partition opposite to their training partition, using the best-validation checkpoint. Metrics include Pearson correlation ($r$) and Spearman correlation ($\rho$).}
\label{tab:oracle_performance}
\resizebox{0.60\textwidth}{!}{%
\begin{tabular}{llcc}
\toprule
\textbf{Model} & \textbf{Cell Line} & \textbf{Pearson $r$} & \textbf{Spearman $\rho$} \\
\midrule
\multirow{3}{*}{Enformer Design}& HepG2 & 0.865 & 0.779 \\
& K562 & 0.862 & 0.744 \\
& SK-N-SH & 0.862 & 0.789 \\
\midrule
\multirow{3}{*}{Enformer Evaluation}& HepG2 & 0.863 & 0.775 \\
& K562 & 0.857 & 0.739 \\
& SK-N-SH & 0.863 & 0.788 \\
\midrule
\multirow{3}{*}{Enformer Full}& HepG2 & 0.918 & 0.829 \\
& K562 & 0.916 & 0.794 \\
& SK-N-SH & 0.916 & 0.835 \\
\midrule
\multirow{3}{*}{Borzoi Design}& HepG2 & 0.865 & 0.781 \\
& K562 & 0.865 & 0.750 \\
& SK-N-SH & 0.865 & 0.788 \\
\midrule
\multirow{3}{*}{Borzoi Evaluation}& HepG2 & 0.864 & 0.779 \\
& K562 & 0.857 & 0.746 \\
& SK-N-SH & 0.857 & 0.784 \\
\bottomrule
\end{tabular}%
}
\end{table}

\begin{table}[h]
\centering
\caption{Validation performance of chromatin accessibility predictors. We report AUROC for the eight immune cell types used in the design benchmark, together with their positive-label prevalence. The final rows summarize the mean across these eight cell types and across all 204 accessibility tracks. Models are evaluated on the partition opposite to their training partition using the best-validation checkpoint.}
\label{tab:oracle_immune_performance}
\resizebox{0.95\textwidth}{!}{%
\begin{tabular}{lcccc}
\toprule
\textbf{Cell Type} & \textbf{Prevalence (\%)} & \textbf{Enformer Design} & \textbf{Enformer Evaluation} & \textbf{Borzoi Evaluation} \\
& & \textbf{AUROC} & \textbf{AUROC} & \textbf{AUROC} \\
\midrule
CD8$^+$ T & 5.24 & 0.925 & 0.923 & 0.922 \\
CD4$^+$ T & 4.43 & 0.931 & 0.928 & 0.927 \\
NKT & 4.04 & 0.934 & 0.934 & 0.934 \\
Naive T & 4.67 & 0.925 & 0.922 & 0.919 \\
Memory B & 4.02 & 0.930 & 0.928 & 0.926 \\
Plasma B & 5.25 & 0.917 & 0.914 & 0.909 \\
Macrophage & 9.23 & 0.907 & 0.905 & 0.903 \\
Mast & 5.22 & 0.918 & 0.918 & 0.912 \\
\midrule
Mean (8 immune types) & 5.26 & 0.923 & 0.922 & 0.919 \\
Mean (all 204 tracks) & -- & 0.900 & 0.898 & 0.894 \\
\bottomrule
\end{tabular}%
}
\end{table}

\subsection{Extended Ablations}
\label{app:extended_ablations}

We provide the complete numerical results for all ablation sweeps discussed in Section~\ref{sec:ablations}. All experiments were conducted on the SK-N-SH enhancer design task, generating 128 sequences per configuration across 3 independent runs.

\textbf{CFG Guidance Scale Sweep.}

Table~\ref{tab:cfg_sweep_full} reports the full metrics for the class-conditioned expansion sweep across guidance scales $\gamma \in \{0.0, 0.5, 1.0, 2.0, 3.0, 4.0, 5.0\}$.

\begin{table}[ht]
\centering
\caption{Full results for the CFG guidance scale sweep during class-conditioned Monte Carlo tree guidance ($M=8$, $B=800$) for SK-N-SH enhancer design. Values are reported as mean $\pm$ std over 3 independent runs.}
\label{tab:cfg_sweep_full}
\resizebox{\textwidth}{!}{%
\begin{tabular}{lcccccc}
\toprule
$\boldsymbol{\gamma}$ & \textbf{MinGap} ($\uparrow$) & \textbf{MinGap (Full MPRA)} ($\uparrow$) & \textbf{Motif Corr.} ($\uparrow$) & \textbf{3-mer Corr.} ($\uparrow$) & \textbf{Shannon Div.} ($\uparrow$) & \textbf{Wall Time (s)} ($\downarrow$) \\
\midrule
0.0 & 5.440 \scriptsize{$\pm 0.038$} & 6.758 \scriptsize{$\pm 0.051$} & \textbf{0.821} \scriptsize{$\pm 0.024$} & 0.909 \scriptsize{$\pm 0.004$} & 1.972 \scriptsize{$\pm 0.001$} & 1021.270 \scriptsize{$\pm 9.646$}\\
0.5 & 5.453 \scriptsize{$\pm 0.048$} & 6.816 \scriptsize{$\pm 0.010$} & 0.803 \scriptsize{$\pm 0.011$} & 0.909 \scriptsize{$\pm 0.005$} & \textbf{1.973} \scriptsize{$\pm 0.001$} & 1712.908 \scriptsize{$\pm 15.843$} \\
1.0 & 5.479 \scriptsize{$\pm 0.057$} & 6.894 \scriptsize{$\pm 0.102$} & 0.809 \scriptsize{$\pm 0.010$} & 0.912 \scriptsize{$\pm 0.001$} & 1.972 \scriptsize{$\pm 0.001$} & 1012.475 \scriptsize{$\pm 3.999$} \\
2.0 & 5.538 \scriptsize{$\pm 0.018$} & 6.893 \scriptsize{$\pm 0.058$} & 0.810 \scriptsize{$\pm 0.003$} & 0.914 \scriptsize{$\pm 0.004$} & 1.972 \scriptsize{$\pm 0.001$} & 1719.055 \scriptsize{$\pm 20.178$} \\
3.0 & 5.529 \scriptsize{$\pm 0.047$} & 6.871 \scriptsize{$\pm 0.034$} & 0.815 \scriptsize{$\pm 0.015$} & 0.918 \scriptsize{$\pm 0.008$} & 1.972 \scriptsize{$\pm 0.001$} & 1718.087 \scriptsize{$\pm 10.862$} \\
4.0 & 5.576 \scriptsize{$\pm 0.100$} & 6.941 \scriptsize{$\pm 0.128$} & 0.810 \scriptsize{$\pm 0.014$} & 0.919 \scriptsize{$\pm 0.005$} & 1.970 \scriptsize{$\pm 0.002$} & 1711.341 \scriptsize{$\pm 7.905$} \\
5.0 & \textbf{5.644} \scriptsize{$\pm 0.032$} & \textbf{7.020} \scriptsize{$\pm 0.046$} & 0.805 \scriptsize{$\pm 0.020$} & \textbf{0.920} \scriptsize{$\pm 0.004$} & 1.970 \scriptsize{$\pm 0.001$} & 1733.893 \scriptsize{$\pm 27.677$} \\
\bottomrule
\end{tabular}%
}
\end{table}

\textbf{Fast Lookahead Pruning Sweep. }

Table~\ref{tab:lookahead_sweep_full} reports the full metrics for the lookahead pool size sweep across $B\in\{2,4,8,16,32,64,128,256,512,1024,2048\}$. The retained-child limit is $M=8$, giving $\min(B,8)$ retained children per expansion. We compare these configurations with standard full rollouts without lookahead (Std.).

\begin{table}[ht]
\centering
\caption{Full results for the fast lookahead pruning sweep during Monte Carlo tree guidance with retained-child limit $M=8$ and guidance scale $\gamma=1.0$ for SK-N-SH enhancer design. Each expansion retains $\min(B,8)$ children. ``Std.'' denotes standard full rollouts without lookahead. Values are reported as mean $\pm$ std over 3 independent runs.}
\label{tab:lookahead_sweep_full}
\resizebox{\textwidth}{!}{%
\begin{tabular}{lcccccc}
\toprule
$\boldsymbol{B}$ & \textbf{MinGap} ($\uparrow$) & \textbf{MinGap (Full MPRA)} ($\uparrow$) & \textbf{Motif Corr.} ($\uparrow$) & \textbf{3-mer Corr.} ($\uparrow$) & \textbf{Shannon Div.} ($\uparrow$) & \textbf{Wall Time (s)} ($\downarrow$) \\
\midrule
Std. & 1.538 \scriptsize{$\pm 0.029$} & 1.658 \scriptsize{$\pm 0.041$} & \textbf{0.858} \scriptsize{$\pm 0.024$} & \textbf{0.947} \scriptsize{$\pm 0.010$} & \textbf{1.980} \scriptsize{$\pm 0.003$} & 1459.713 \scriptsize{$\pm 4.460$} \\
2 & 1.554 \scriptsize{$\pm 0.070$} & 1.692 \scriptsize{$\pm 0.088$} & 0.811 \scriptsize{$\pm 0.021$} & 0.866 \scriptsize{$\pm 0.007$} & 1.967 \scriptsize{$\pm 0.003$} & 221.196 \scriptsize{$\pm 5.063$}\\
4 & 2.069 \scriptsize{$\pm 0.069$} & 2.319 \scriptsize{$\pm 0.100$} & 0.797 \scriptsize{$\pm 0.020$} & 0.899 \scriptsize{$\pm 0.019$} & 1.979 \scriptsize{$\pm 0.003$} & 217.319 \scriptsize{$\pm 8.547$} \\
8 & 2.772 \scriptsize{$\pm 0.035$} & 3.200 \scriptsize{$\pm 0.071$} & 0.777 \scriptsize{$\pm 0.025$} & 0.908 \scriptsize{$\pm 0.008$} & 1.979 \scriptsize{$\pm 0.002$} & 225.092 \scriptsize{$\pm 3.689$} \\
16 & 3.501 \scriptsize{$\pm 0.059$} & 4.067 \scriptsize{$\pm 0.130$} & 0.801 \scriptsize{$\pm 0.043$} & 0.914 \scriptsize{$\pm 0.008$} & \textbf{1.980} \scriptsize{$\pm 0.001$} & 229.305 \scriptsize{$\pm 8.031$} \\
32 & 3.934 \scriptsize{$\pm 0.061$} & 4.746 \scriptsize{$\pm 0.040$} & 0.809 \scriptsize{$\pm 0.004$} & 0.917 \scriptsize{$\pm 0.009$} & \textbf{1.980} \scriptsize{$\pm 0.001$} & 231.643 \scriptsize{$\pm 5.718$} \\
64 & 4.275 \scriptsize{$\pm 0.095$} & 5.291 \scriptsize{$\pm 0.106$} & 0.814 \scriptsize{$\pm 0.007$} & 0.920 \scriptsize{$\pm 0.008$} & 1.978 \scriptsize{$\pm 0.001$} & 243.243 \scriptsize{$\pm 7.647$} \\
128 & 4.785 \scriptsize{$\pm 0.057$} & 5.960 \scriptsize{$\pm 0.097$} & 0.815 \scriptsize{$\pm 0.006$} & 0.915 \scriptsize{$\pm 0.002$} & 1.977 \scriptsize{$\pm 0.001$} & 257.311 \scriptsize{$\pm 3.575$} \\
256 & 5.079 \scriptsize{$\pm 0.056$} & 6.266 \scriptsize{$\pm 0.112$} & 0.828 \scriptsize{$\pm 0.016$} & 0.924 \scriptsize{$\pm 0.001$} & 1.975 \scriptsize{$\pm 0.001$} & 368.128 \scriptsize{$\pm 0.834$} \\
512 & 5.382 \scriptsize{$\pm 0.010$} & 6.743 \scriptsize{$\pm 0.053$} & 0.825 \scriptsize{$\pm 0.007$} & 0.913 \scriptsize{$\pm 0.006$} & 1.973 \scriptsize{$\pm 0.001$} & 676.850 \scriptsize{$\pm 9.089$} \\
1024 & 5.554 \scriptsize{$\pm 0.016$} & 6.958 \scriptsize{$\pm 0.043$} & 0.792 \scriptsize{$\pm 0.033$} & 0.909 \scriptsize{$\pm 0.010$} & 1.973 \scriptsize{$\pm 0.001$} & 1309.609 \scriptsize{$\pm 9.239$} \\
2048 & \textbf{5.590} \scriptsize{$\pm 0.043$} & \textbf{6.965} \scriptsize{$\pm 0.124$} & 0.819 \scriptsize{$\pm 0.017$} & 0.904 \scriptsize{$\pm 0.009$} & 1.972 \scriptsize{$\pm 0.002$} & 2932.594 \scriptsize{$\pm 3.048$} \\
\bottomrule
\end{tabular}%
}
\end{table}

\textbf{Cross-Method Evaluation under Alternative Specificity Metrics.}
\label{app:alternative_specificity}

To assess whether the specificity comparisons depend on the choice of evaluation metric, we evaluated the generated sequences from all seven methods using three alternative scores under the Enformer Evaluation model. MinGap subtracts the maximum off-target activity from the target activity, LogSumExp replaces this maximum with a smooth aggregation, and Mean Difference subtracts the mean off-target activity. The generated sequences are held fixed across these evaluations. This analysis complements the reward-formulation ablation in Table~\ref{tab:reward_ablation}, which changes the objective used during DNA-CRAFT optimization.

Table~\ref{tab:alternative_specificity} shows that the leading method within each cell line is consistent across all three metrics. Ctrl-DNA achieves the highest mean specificity scores on HepG2 and K562, while DNA-CRAFT achieves the highest scores on SK-N-SH. Thus, these leading specificity comparisons are preserved under alternative off-target aggregations, although individual method rankings can change. Together with the naturalness results in Table~\ref{tab:human_enhancer_extended}, this supports evaluating the methods through their specificity--naturalness trade-off rather than a single specificity score.

\begin{table}[h]
\centering
\caption{Cross-method specificity evaluation across three human cell lines. The same generated sequences are evaluated using hard MinGap, smooth LogSumExp, and Mean Difference. Higher values are better for all three metrics. Values are reported as mean $\pm$ standard deviation over three independent runs, with the highest mean in each row marked in \textbf{bold}.}
\label{tab:alternative_specificity}
\resizebox{\textwidth}{!}{%
\begin{tabular}{llccccccc}
\toprule
\textbf{Cell Line} & \textbf{Metric} & \textbf{SMC} & \textbf{TDS} & \textbf{D3} & \textbf{DRAKES} & \textbf{Ctrl-DNA} & \textbf{Ledidi} & \textbf{DNA-CRAFT (Ours)} \\
\midrule
\multirow{3}{*}{\textbf{HepG2}} & MinGap $\uparrow$ & 0.989 \scriptsize{$\pm 1.717$} & 0.880 \scriptsize{$\pm 1.053$} & -0.075 \scriptsize{$\pm 0.021$} & -1.314 \scriptsize{$\pm 0.024$} & \textbf{7.464} \scriptsize{$\pm 0.270$} & 6.712 \scriptsize{$\pm 0.270$} & 7.271 \scriptsize{$\pm 0.142$} \\
 & LogSumExp $\uparrow$ & 0.432 \scriptsize{$\pm 1.669$} & 0.332 \scriptsize{$\pm 1.059$} & -0.616 \scriptsize{$\pm 0.020$} & -1.811 \scriptsize{$\pm 0.049$} & \textbf{7.001} \scriptsize{$\pm 0.298$} & 6.124 \scriptsize{$\pm 0.262$} & 6.737 \scriptsize{$\pm 0.146$} \\
 & Mean Diff. $\uparrow$ & 1.144 \scriptsize{$\pm 1.660$} & 1.046 \scriptsize{$\pm 1.056$} & 0.110 \scriptsize{$\pm 0.019$} & -1.072 \scriptsize{$\pm 0.061$} & \textbf{7.744} \scriptsize{$\pm 0.303$} & 6.828 \scriptsize{$\pm 0.261$} & 7.453 \scriptsize{$\pm 0.148$} \\
\midrule
\multirow{3}{*}{\textbf{K562}} & MinGap $\uparrow$ & 1.812 \scriptsize{$\pm 0.768$} & 1.735 \scriptsize{$\pm 0.660$} & 0.003 \scriptsize{$\pm 0.070$} & -0.033 \scriptsize{$\pm 0.032$} & \textbf{8.490} \scriptsize{$\pm 0.153$} & 8.004 \scriptsize{$\pm 0.078$} & 7.777 \scriptsize{$\pm 0.054$} \\
 & LogSumExp $\uparrow$ & 1.248 \scriptsize{$\pm 0.758$} & 1.224 \scriptsize{$\pm 0.632$} & -0.566 \scriptsize{$\pm 0.078$} & -0.457 \scriptsize{$\pm 0.024$} & \textbf{8.221} \scriptsize{$\pm 0.136$} & 7.644 \scriptsize{$\pm 0.077$} & 7.410 \scriptsize{$\pm 0.056$} \\
 & Mean Diff. $\uparrow$ & 1.957 \scriptsize{$\pm 0.754$} & 1.954 \scriptsize{$\pm 0.624$} & 0.155 \scriptsize{$\pm 0.081$} & 0.333 \scriptsize{$\pm 0.018$} & \textbf{9.090} \scriptsize{$\pm 0.119$} & 8.433 \scriptsize{$\pm 0.075$} & 8.201 \scriptsize{$\pm 0.056$} \\
\midrule
\multirow{3}{*}{\textbf{SK-N-SH}} & MinGap $\uparrow$ & 0.252 \scriptsize{$\pm 0.161$} & 1.170 \scriptsize{$\pm 0.715$} & -0.111 \scriptsize{$\pm 0.014$} & -0.082 \scriptsize{$\pm 0.040$} & 3.988 \scriptsize{$\pm 0.856$} & 3.202 \scriptsize{$\pm 0.258$} & \textbf{5.479} \scriptsize{$\pm 0.057$} \\
 & LogSumExp $\uparrow$ & -0.366 \scriptsize{$\pm 0.178$} & 0.624 \scriptsize{$\pm 0.717$} & -0.683 \scriptsize{$\pm 0.013$} & -0.369 \scriptsize{$\pm 0.042$} & 3.717 \scriptsize{$\pm 0.910$} & 2.770 \scriptsize{$\pm 0.256$} & \textbf{5.222} \scriptsize{$\pm 0.050$} \\
 & Mean Diff. $\uparrow$ & 0.337 \scriptsize{$\pm 0.186$} & 1.339 \scriptsize{$\pm 0.715$} & 0.037 \scriptsize{$\pm 0.019$} & 0.496 \scriptsize{$\pm 0.045$} & 4.706 \scriptsize{$\pm 0.990$} & 3.549 \scriptsize{$\pm 0.253$} & \textbf{6.134} \scriptsize{$\pm 0.035$} \\
\bottomrule
\end{tabular}%
}
\end{table}

\subsection{Tree Search vs.\ Best-of-$B$ Greedy Baseline}
\label{app:tree_vs_greedy}

\begin{table}[h]
\centering
\caption{Tree-free best-of-$B$ greedy baseline with identical model inference parameters. MinGap ($\uparrow$) is measured under the held-out Evaluation oracle.}
\label{tab:tree_vs_greedy}
\resizebox{0.7\textwidth}{!}{%
\begin{tabular}{lccc}
\toprule
\textbf{Method} & \textbf{HepG2} & \textbf{K562} & \textbf{SK-N-SH} \\
\midrule
Best-of-$B$ Greedy & $6.87$\scriptsize{$\pm 0.18$} & $6.94$\scriptsize{$\pm 0.02$} & $5.36$\scriptsize{$\pm 0.08$} \\
DNA-CRAFT (MCTD)   & $\textbf{7.27}$\scriptsize{$\pm 0.14$} & $\textbf{7.78}$\scriptsize{$\pm 0.05$} & $\textbf{5.48}$\scriptsize{$\pm 0.06$} \\
\bottomrule
\end{tabular}%
}
\end{table}

To assess the contribution of the complete tree-search procedure, we compare DNA-CRAFT with a tree-free best-of-$B$ greedy baseline, with both methods configured for a maximum of 32 candidate-pool evaluations of $B=800$ sequences. For each run, we report the best sequence from the final batch of 800 fully denoised candidates for best-of-$B$ sampling. Table~\ref{tab:tree_vs_greedy} shows higher mean specificity for DNA-CRAFT under the matched maximum candidate budget, with the observed differences reflecting the combined search and final-selection procedures.

\subsection{Tree Depth Statistics}
\label{app:tree_depth}

We recorded the maximum depth reached across 128 independent SK-N-SH searches with a budget of 32 iterations. Of these, 122 searches (95.3\%) reached depth 32, while the remaining six reached depths 4, 4, 20, 21, 25, and 26. Because each iteration expands at most one node, the depth-32 searches allocate their expansions along a single path. The remaining searches show that the iteration budget is not always used to increase the maximum depth. These observations characterize search behavior at the default budget, while the comparison in Table~\ref{tab:tree_vs_greedy} evaluates the performance of the complete search procedures.

\subsection{Evaluation of the One-Step MAP Estimator}
\label{app:map_validation}

The one-step MAP estimator (Equation~\ref{eq:x0_map}) approximates sequence quality at an intermediate diffusion state using a deterministic completion. We evaluated this approximation using child candidates collected from 32 SK-N-SH tree searches at five denoising steps. Each partially masked sequence was completed once using the MAP estimator and eight times using stochastic reverse-diffusion rollouts. We compared its MAP MinGap score with the average MinGap across the eight rollouts. Table~\ref{tab:map_validation} reports the mean signed difference, its standard deviation across candidates, and the Pearson correlation between the two scores. The step indices count denoising updates from the fully masked state and are distinct from the continuous diffusion time $t$. Steps 2 and 8 show near-zero Pearson correlation, while the measured correlation increases at later steps, reaching $r=0.44$ at step 30.

\begin{table}[h]
\centering
\caption{\textbf{One-step MAP estimator accuracy across denoising steps for SK-N-SH.} MAP scores are compared with the average scores from eight stochastic rollouts per candidate. Shown are the number of candidates, the mean signed difference (MAP $-$ rollout average) $\pm$ standard deviation across candidates, and the Pearson correlation $r$.}
\label{tab:map_validation}
\resizebox{0.75\textwidth}{!}{%
\begin{tabular}{cccc}
\toprule
\textbf{Denoising Step}& \textbf{Sequences} & \textbf{MAP $-$ Rollout (mean $\pm$ std)} & \textbf{Pearson $r$} \\
\midrule
2  & 2592 & $+0.01$\scriptsize{$\pm 0.83$} & $0.00$ \\
8  & 1088 & $+0.14$\scriptsize{$\pm 0.88$} & $0.03$ \\
16 & 960  & $+0.35$\scriptsize{$\pm 0.73$} & $0.33$ \\
24 & 896  & $+0.31$\scriptsize{$\pm 0.68$} & $0.41$ \\
30 & 672  & $+0.19$\scriptsize{$\pm 0.96$} & $0.44$ \\
\bottomrule
\end{tabular}%
}
\end{table}

To assess the utility of the completion rule for pruning, we repeated the lookahead search with a candidate pool of $B=800$, changing only the score used for candidate selection. One configuration scored one-step MAP completions, while the other scored completions obtained by filling masked positions with random nucleotides. The MAP configuration achieved a best MinGap of $6.17 \pm 1.12$, compared with $3.75 \pm 1.30$ for random completion. This comparison supports the usefulness of the MAP-based pruning score despite its weak association with rollout scores at early denoising steps.


\clearpage

\end{document}

%% file: math_commands.tex
\usepackage{amsmath,amsfonts,bm}
\usepackage{hyperref}       %

\def\eqref#1{equation~\ref{#1}}

\def\1{\bm{1}}

\DeclareMathAlphabet{\mathsfit}{\encodingdefault}{\sfdefault}{m}{sl}
\SetMathAlphabet{\mathsfit}{bold}{\encodingdefault}{\sfdefault}{bx}{n}

%% file: neurips2026_conference.bib
@inproceedings{GFlowNet,
  title={Biological sequence design with gflownets},
  author={Jain, Moksh and Bengio, Emmanuel and Hernandez-Garcia, Alex and Rector-Brooks, Jarrid and Dossou, Bonaventure FP and Ekbote, Chanakya Ajit and Fu, Jie and Zhang, Tianyu and Kilgour, Michael and Zhang, Dinghuai and others},
  booktitle={International Conference on Machine Learning},
  pages={9786--9801},
  year={2022},
  organization={PMLR}
}

@article{linder2021fast,
  title={Fast activation maximization for molecular sequence design},
  author={Linder, Johannes and Seelig, Georg},
  journal={BMC bioinformatics},
  volume={22},
  pages={1--20},
  year={2021},
  publisher={Springer}
}

@article{vaishnav2022evolution,
  title={The evolution, evolvability and engineering of gene regulatory DNA},
  author={Vaishnav, Eeshit Dhaval and de Boer, Carl G and Molinet, Jennifer and Yassour, Moran and Fan, Lin and Adiconis, Xian and Thompson, Dawn A and Levin, Joshua Z and Cubillos, Francisco A and Regev, Aviv},
  journal={Nature},
  volume={603},
  number={7901},
  pages={455--463},
  year={2022},
  publisher={Nature Publishing Group UK London}
}

@inproceedings{avdeyev2023dirichlet,
  title={Dirichlet diffusion score model for biological sequence generation},
  author={Avdeyev, Pavel and Shi, Chenlai and Tan, Yuhao and Dudnyk, Kseniia and Zhou, Jian},
  booktitle={International Conference on Machine Learning},
  pages={1276--1301},
  year={2023},
  organization={PMLR}
}

@article{gosai2024machine,
  title={Machine-guided design of cell-type-targeting cis-regulatory elements},
  author={Gosai, Sager J and Castro, Rodrigo I and Fuentes, Natalia and Butts, John C and Mouri, Kousuke and Alasoadura, Michael and Kales, Susan and Nguyen, Thanh Thanh L and Noche, Ramil R and Rao, Arya S and others},
  journal={Nature},
  pages={1--10},
  year={2024},
  publisher={Nature Publishing Group UK London}
}

@article{reddy2024designing,
  title={Designing Cell-Type-Specific Promoter Sequences Using Conservative Model-Based Optimization},
  author={Reddy, Aniketh Janardhan and Geng, Xinyang and Herschl, Michael and Kolli, Sathvik and Kumar, Aviral and Hsu, Patrick and Levine, Sergey and Ioannidis, Nilah},
  journal={Advances in Neural Information Processing Systems},
  volume={37},
  pages={93033--93059},
  year={2024}
}

@article{de2020deciphering,
  title={Deciphering eukaryotic gene-regulatory logic with 100 million random promoters},
  author={de Boer, Carl G and Vaishnav, Eeshit Dhaval and Sadeh, Ronen and Abeyta, Esteban Luis and Friedman, Nir and Regev, Aviv},
  journal={Nature biotechnology},
  volume={38},
  number={1},
  pages={56--65},
  year={2020},
  publisher={Nature Publishing Group US New York}
}

@article{reglm,
  title={Designing realistic regulatory DNA with autoregressive language models},
  author={Lal, Avantika and Garfield, David and Biancalani, Tommaso and Eraslan, Gokcen},
  journal={Genome Research},
  volume={34},
  number={9},
  pages={1411--1420},
  year={2024},
  publisher={Cold Spring Harbor Lab}
}

@article{enformer,
  title={Effective gene expression prediction from sequence by integrating long-range interactions},
  author={Avsec, {\v{Z}}iga and Agarwal, Vikram and Visentin, Daniel and Ledsam, Joseph R and Grabska-Barwinska, Agnieszka and Taylor, Kyle R and Assael, Yannis and Jumper, John and Kohli, Pushmeet and Kelley, David R},
  journal={Nature methods},
  volume={18},
  number={10},
  pages={1196--1203},
  year={2021},
  publisher={Nature Publishing Group US New York}
}

@article{nguyen2023hyenadna,
  title={Hyenadna: Long-range genomic sequence modeling at single nucleotide resolution},
  author={Nguyen, Eric and Poli, Michael and Faizi, Marjan and Thomas, Armin and Wornow, Michael and Birch-Sykes, Callum and Massaroli, Stefano and Patel, Aman and Rabideau, Clayton and Bengio, Yoshua and others},
  journal={Advances in neural information processing systems},
  volume={36},
  pages={43177--43201},
  year={2023}
}

@article{sinai2020adalead,
  title={Adalead: A simple and robust adaptive greedy search algorithm for sequence design},
  author={Sinai, Sam and Wang, Richard and Whatley, Alexander and Slocum, Stewart and Locane, Elina and Kelsic, Eric D},
  journal={arXiv preprint arXiv:2010.02141},
  year={2020}
}

@article{moore_expanded_2026,
	title = {An expanded registry of candidate cis-regulatory elements},
	copyright = {2026 The Author(s)},
	issn = {1476-4687},
	url = {https://www.nature.com/articles/s41586-025-09909-9},
	doi = {10.1038/s41586-025-09909-9},
	language = {en},
	urldate = {2026-01-21},
	journal = {Nature},
	author = {Moore, Jill E. and Pratt, Henry E. and Fan, Kaili and Phalke, Nishigandha and Fisher, Jonathan and Elhajjajy, Shaimae I. and Andrews, Gregory and Gao, Mingshi and Shedd, Nicole and Fu, Yu and Lacadie, Matthew C. and Meza, Jair and Khandpekar, Mansi and Ganna, Mohit and Choudhury, Eva and Swofford, Ross and Phan, Huong and Ramirez, Christian C. and Campbell, Maxwell and Likhite, Mary and Farrell, Nina P. and Weimer, Annika K. and Pampari, Anusri and Ramalingam, Vivekanandan and Reese, Fairlie and Borsari, Beatrice and Yu, Xuezhu and Wattenberg, Eve and Ruiz-Romero, Marina and Razavi-Mohseni, Milad and Xu, Jinrui and Galeev, Timur and Colubri, Andres and Beer, Michael A. and Guigó, Roderic and Gerstein, Mark B. and Engreitz, Jesse M. and Ljungman, Mats and Reddy, Timothy E. and Snyder, Michael P. and Epstein, Charles B. and Gaskell, Elizabeth and Bernstein, Bradley E. and Dickel, Diane E. and Visel, Axel and Pennacchio, Len A. and Mortazavi, Ali and Kundaje, Anshul and Weng, Zhiping},
	month = jan,
	year = {2026},
	note = {Publisher: Nature Publishing Group},
	pages = {1--10},
}

@article{dasilva_designing_2026,
	title = {Designing synthetic regulatory elements using the generative {AI} framework {DNA}-{Diffusion}},
	volume = {58},
	copyright = {2025 The Author(s), under exclusive licence to Springer Nature America, Inc.},
	issn = {1546-1718},
	url = {https://www.nature.com/articles/s41588-025-02441-6},
	doi = {10.1038/s41588-025-02441-6},
	language = {en},
	number = {1},
	urldate = {2026-01-21},
	journal = {Nature Genetics},
	author = {DaSilva, Lucas Ferreira and Senan, Simon and Kribelbauer-Swietek, Judith F. and Patel, Zain Munir and Louis, Lithin Karmel and Reddy, Aniketh Janardhan and Gabbita, Sameer and Rosen, Jonathan D. and Nussbaum, Zach and Córdova, C{\'e}sar Miguel Valdez and Wenteler, Aaron and Weber, Noah and Tunjic, Tin M. and Mansoldo, Martino and Khan, Talha Ahmad and Hwang, Gue-Ho and Gardeux, Vincent and Humphreys, David T. and Smith, Cameron and Bejan, Matei and Bromley, Peter and Connell, Will and Deplancke, Bart and Love, Michael I. and Wong, Emily S. and Meuleman, Wouter and Pinello, Luca},
	month = jan,
	year = {2026},
	note = {Publisher: Nature Publishing Group},
	pages = {180--194},
}

@misc{tang_peptune_2024,
	title = {{PepTune}: {De} {Novo} {Generation} of {Therapeutic} {Peptides} with {Multi}-{Objective}-{Guided} {Discrete} {Diffusion}},
	shorttitle = {{PepTune}},
	url = {https://arxiv.org/abs/2412.17780v4},
	language = {en},
	urldate = {2026-01-21},
	journal = {arXiv.org},
	author = {Tang, Sophia and Zhang, Yinuo and Chatterjee, Pranam},
	month = dec,
	year = {2024},
}

@article{zhang_single-cell_2021,
	title = {A single-cell atlas of chromatin accessibility in the human genome},
	volume = {184},
	issn = {0092-8674, 1097-4172},
	url = {https://www.cell.com/cell/abstract/S0092-8674(21)01279-4},
	doi = {10.1016/j.cell.2021.10.024},
	language = {English},
	number = {24},
	urldate = {2026-01-21},
	journal = {Cell},
	author = {Zhang, Kai and Hocker, James D. and Miller, Michael and Hou, Xiaomeng and Chiou, Joshua and Poirion, Olivier B. and Qiu, Yunjiang and Li, Yang E. and Gaulton, Kyle J. and Wang, Allen and Preissl, Sebastian and Ren, Bing},
	month = nov,
	year = {2021},
	pmid = {34774128},
	note = {Publisher: Elsevier},
	pages = {5985--6001.e19},
}

@misc{wang_fine-tuning_2025,
	title = {Fine-{Tuning} {Discrete} {Diffusion} {Models} via {Reward} {Optimization} with {Applications} to {DNA} and {Protein} {Design}},
	url = {http://arxiv.org/abs/2410.13643},
	doi = {10.48550/arXiv.2410.13643},
	urldate = {2026-01-21},
	publisher = {arXiv},
	author = {Wang, Chenyu and Uehara, Masatoshi and He, Yichun and Wang, Amy and Biancalani, Tommaso and Lal, Avantika and Jaakkola, Tommi and Levine, Sergey and Wang, Hanchen and Regev, Aviv},
	month = mar,
	year = {2025},
	note = {arXiv:2410.13643 [cs]},
}

@book{laarhoven_simulated_1987,
	title = {Simulated {Annealing}: {Theory} and {Applications}},
	isbn = {978-90-277-2513-4},
	shorttitle = {Simulated {Annealing}},
	language = {en},
	publisher = {Springer Science \& Business Media},
	author = {Laarhoven, P. J. van and Aarts, E. H.},
	month = jun,
	year = {1987},
}

@misc{schreiber_programmatic_2025,
	title = {Programmatic design and editing of cis-regulatory elements},
	copyright = {© 2025, Posted by openRxiv. This pre-print is available under a Creative Commons License (Attribution 4.0 International), CC BY 4.0, as described at http://creativecommons.org/licenses/by/4.0/},
	url = {https://www.biorxiv.org/content/10.1101/2025.04.22.650035v2},
	doi = {10.1101/2025.04.22.650035},
	language = {en},
	urldate = {2026-01-21},
	publisher = {bioRxiv},
	author = {Schreiber, Jacob and Lorbeer, Franziska Katharina and Heinzl, Monika and Reiter, Franziska and Rafanel, Baptiste and Lu, Yang Young and Stark, Alexander and Noble, William Stafford},
	month = dec,
	year = {2025},
	note = {ISSN: 2692-8205
Pages: 2025.04.22.650035
Section: New Results},
}

@inproceedings{angermueller_model-based_2019,
	title = {Model-based reinforcement learning for biological sequence design},
	url = {https://openreview.net/forum?fileGuid=3xgr169o12oUrbxS&id=HklxbgBKvr&ref=https%3A%2F%2Fgithubhelp.com},
	language = {en},
	urldate = {2026-01-21},
	author = {Angermueller, Christof and Dohan, David and Belanger, David and Deshpande, Ramya and Murphy, Kevin and Colwell, Lucy},
	month = sep,
	year = {2019},
}

@misc{sarkar_designing_2025,
	title = {Designing {DNA} {With} {Tunable} {Regulatory} {Activity} {Using} {Score}-{Entropy} {Discrete} {Diffusion}},
	copyright = {© 2025, Posted by Cold Spring Harbor Laboratory. This pre-print is available under a Creative Commons License (Attribution-NonCommercial-NoDerivs 4.0 International), CC BY-NC-ND 4.0, as described at http://creativecommons.org/licenses/by-nc-nd/4.0/},
	url = {https://www.biorxiv.org/content/10.1101/2024.05.23.595630v2},
	doi = {10.1101/2024.05.23.595630},
	language = {en},
	urldate = {2026-01-21},
	publisher = {bioRxiv},
	author = {Sarkar, Anirban and Kang, Yijie and Somia, Nirali and Puccetti, Pablo Mantilla and Zhou, Jessica and Nagai, Masayuki and Tang, Ziqi and Zhao, Chris and Koo, Peter K.},
	month = may,
	year = {2025},
	note = {Pages: 2024.05.23.595630
Section: New Results},
}

@article{mitra_single-cell_2024,
	title = {Single-cell multi-ome regression models identify functional and disease-associated enhancers and enable chromatin potential analysis},
	volume = {56},
	copyright = {2024 The Author(s)},
	issn = {1546-1718},
	url = {https://www.nature.com/articles/s41588-024-01689-8},
	doi = {10.1038/s41588-024-01689-8},
	language = {en},
	number = {4},
	urldate = {2026-01-21},
	journal = {Nature Genetics},
	author = {Mitra, Sneha and Malik, Rohan and Wong, Wilfred and Rahman, Afsana and Hartemink, Alexander J. and Pritykin, Yuri and Dey, Kushal K. and Leslie, Christina S.},
	month = apr,
	year = {2024},
	note = {Publisher: Nature Publishing Group},
	pages = {627--636},
}

@misc{chen_ctrl-dna_2025,
	title = {Ctrl-{DNA}: {Controllable} {Cell}-{Type}-{Specific} {Regulatory} {DNA} {Design} via {Constrained} {RL}},
	shorttitle = {Ctrl-{DNA}},
	url = {http://arxiv.org/abs/2505.20578},
	doi = {10.48550/arXiv.2505.20578},
	urldate = {2026-01-21},
	publisher = {arXiv},
	author = {Chen, Xingyu and Ma, Shihao and Lin, Runsheng and Lin, Jiecong and Wang, Bo},
	month = may,
	year = {2025},
	note = {arXiv:2505.20578 [cs]},
}

@misc{wu_practical_2024,
	title = {Practical and {Asymptotically} {Exact} {Conditional} {Sampling} in {Diffusion} {Models}},
	url = {http://arxiv.org/abs/2306.17775},
	doi = {10.48550/arXiv.2306.17775},
	urldate = {2026-01-21},
	publisher = {arXiv},
	author = {Wu, Luhuan and Trippe, Brian L. and Naesseth, Christian A. and Blei, David M. and Cunningham, John P.},
	month = nov,
	year = {2024},
	note = {arXiv:2306.17775 [stat]},
}

@misc{nisonoff_unlocking_2025,
	title = {Unlocking {Guidance} for {Discrete} {State}-{Space} {Diffusion} and {Flow} {Models}},
	url = {http://arxiv.org/abs/2406.01572},
	doi = {10.48550/arXiv.2406.01572},
	urldate = {2026-01-21},
	publisher = {arXiv},
	author = {Nisonoff, Hunter and Xiong, Junhao and Allenspach, Stephan and Listgarten, Jennifer},
	month = mar,
	year = {2025},
	note = {arXiv:2406.01572 [cs]},
}

@misc{phillips_particle_2024,
	title = {Particle {Denoising} {Diffusion} {Sampler}},
	url = {http://arxiv.org/abs/2402.06320},
	doi = {10.48550/arXiv.2402.06320},
	urldate = {2026-01-21},
	publisher = {arXiv},
	author = {Phillips, Angus and Dau, Hai-Dang and Hutchinson, Michael John and Bortoli, Valentin De and Deligiannidis, George and Doucet, Arnaud},
	month = jun,
	year = {2024},
	note = {arXiv:2402.06320 [stat]},
}

@misc{li_derivative-free_2024,
	title = {Derivative-{Free} {Guidance} in {Continuous} and {Discrete} {Diffusion} {Models} with {Soft} {Value}-{Based} {Decoding}},
	url = {http://arxiv.org/abs/2408.08252},
	doi = {10.48550/arXiv.2408.08252},
	urldate = {2026-01-21},
	publisher = {arXiv},
	author = {Li, Xiner and Zhao, Yulai and Wang, Chenyu and Scalia, Gabriele and Eraslan, Gokcen and Nair, Surag and Biancalani, Tommaso and Ji, Shuiwang and Regev, Aviv and Levine, Sergey and Uehara, Masatoshi},
	month = oct,
	year = {2024},
	note = {arXiv:2408.08252 [cs]},
}

@misc{sahoo_simple_2024,
	title = {Simple and {Effective} {Masked} {Diffusion} {Language} {Models}},
	url = {http://arxiv.org/abs/2406.07524},
	doi = {10.48550/arXiv.2406.07524},
	urldate = {2026-01-21},
	publisher = {arXiv},
	author = {Sahoo, Subham Sekhar and Arriola, Marianne and Schiff, Yair and Gokaslan, Aaron and Marroquin, Edgar and Chiu, Justin T. and Rush, Alexander and Kuleshov, Volodymyr},
	month = nov,
	year = {2024},
	note = {arXiv:2406.07524 [cs]},
}

@misc{schiff_simple_2025,
	title = {Simple {Guidance} {Mechanisms} for {Discrete} {Diffusion} {Models}},
	url = {http://arxiv.org/abs/2412.10193},
	doi = {10.48550/arXiv.2412.10193},
	urldate = {2026-01-21},
	publisher = {arXiv},
	author = {Schiff, Yair and Sahoo, Subham Sekhar and Phung, Hao and Wang, Guanghan and Boshar, Sam and Dalla-torre, Hugo and Almeida, Bernardo P. de and Rush, Alexander and Pierrot, Thomas and Kuleshov, Volodymyr},
	month = may,
	year = {2025},
	note = {arXiv:2412.10193 [cs]},
}

@misc{schiff_caduceus_2024,
	title = {Caduceus: {Bi}-{Directional} {Equivariant} {Long}-{Range} {DNA} {Sequence} {Modeling}},
	shorttitle = {Caduceus},
	url = {http://arxiv.org/abs/2403.03234},
	doi = {10.48550/arXiv.2403.03234},
	urldate = {2026-01-21},
	publisher = {arXiv},
	author = {Schiff, Yair and Kao, Chia-Hsiang and Gokaslan, Aaron and Dao, Tri and Gu, Albert and Kuleshov, Volodymyr},
	month = jun,
	year = {2024},
	note = {arXiv:2403.03234 [q-bio]},
}

@article{linder_predicting_2025,
	title = {Predicting {RNA}-seq coverage from {DNA} sequence as a unifying model of gene regulation},
	volume = {57},
	copyright = {2025 The Author(s)},
	issn = {1546-1718},
	url = {https://www.nature.com/articles/s41588-024-02053-6},
	doi = {10.1038/s41588-024-02053-6},
	language = {en},
	number = {4},
	urldate = {2026-01-21},
	journal = {Nature Genetics},
	author = {Linder, Johannes and Srivastava, Divyanshi and Yuan, Han and Agarwal, Vikram and Kelley, David R.},
	month = apr,
	year = {2025},
	note = {Publisher: Nature Publishing Group},
	pages = {949--961},
}

@article{rauluseviciute_jaspar_2024,
	title = {{JASPAR} 2024: 20th anniversary of the open-access database of transcription factor binding profiles},
	volume = {52},
	issn = {0305-1048},
	shorttitle = {{JASPAR} 2024},
	url = {https://doi.org/10.1093/nar/gkad1059},
	doi = {10.1093/nar/gkad1059},
	number = {D1},
	urldate = {2026-01-21},
	journal = {Nucleic Acids Research},
	author = {Rauluseviciute, Ieva and Riudavets-Puig, Rafael and Blanc-Mathieu, Romain and Castro-Mondragon, Jaime A and Ferenc, Katalin and Kumar, Vipin and Lemma, Roza Berhanu and Lucas, J{\'e}r{\'e}my and Ch{\'e}neby, Jeanne and Baranasic, Damir and Khan, Aziz and Fornes, Oriol and Gundersen, Sveinung and Johansen, Morten and Hovig, Eivind and Lenhard, Boris and Sandelin, Albin and Wasserman, Wyeth W and Parcy, Fran${\c{c}}$ois and Mathelier, Anthony},
	month = jan,
	year = {2024},
	pages = {D174--D182},
}

@article{grant_fimo_2011,
	title = {{FIMO}: scanning for occurrences of a given motif},
	volume = {27},
	issn = {1367-4803},
	shorttitle = {{FIMO}},
	url = {https://doi.org/10.1093/bioinformatics/btr064},
	doi = {10.1093/bioinformatics/btr064},
	number = {7},
	urldate = {2026-01-21},
	journal = {Bioinformatics},
	author = {Grant, Charles E. and Bailey, Timothy L. and Noble, William Stafford},
	month = apr,
	year = {2011},
	pages = {1017--1018},
}

@article{yang_enhancer_2025,
	title = {Enhancer reprogramming: critical roles in cancer and promising therapeutic strategies},
	volume = {11},
	copyright = {2025 The Author(s)},
	issn = {2058-7716},
	shorttitle = {Enhancer reprogramming},
	url = {https://www.nature.com/articles/s41420-025-02366-3},
	doi = {10.1038/s41420-025-02366-3},
	language = {en},
	number = {1},
	urldate = {2026-01-21},
	journal = {Cell Death Discovery},
	author = {Yang, Jinshou and Zhou, Feihan and Luo, Xiyuan and Fang, Yuan and Wang, Xing and Liu, Xiaohong and Xiao, Ruiling and Jiang, Decheng and Tang, Yuemeng and Yang, Gang and You, Lei and Zhao, Yupei},
	month = mar,
	year = {2025},
	note = {Publisher: Nature Publishing Group},
	pages = {84},
}

@misc{lou2024discretediffusionmodelingestimating,
      title={Discrete Diffusion Modeling by Estimating the Ratios of the Data Distribution}, 
      author={Aaron Lou and Chenlin Meng and Stefano Ermon},
      year={2024},
      eprint={2310.16834},
      archivePrefix={arXiv},
      primaryClass={stat.ML},
      url={https://arxiv.org/abs/2310.16834}, 
}

@misc{uehara2025inferencetimealignmentdiffusionmodels,
      title={Inference-Time Alignment in Diffusion Models with Reward-Guided Generation: Tutorial and Review}, 
      author={Masatoshi Uehara and Yulai Zhao and Chenyu Wang and Xiner Li and Aviv Regev and Sergey Levine and Tommaso Biancalani},
      year={2025},
      eprint={2501.09685},
      archivePrefix={arXiv},
      primaryClass={cs.AI},
      url={https://arxiv.org/abs/2501.09685}, 
}

@article{lal_grelu_2025,
	title = {{gReLU}: a comprehensive framework for {DNA} sequence modeling and design},
	volume = {22},
	copyright = {2025 The Author(s)},
	issn = {1548-7105},
	shorttitle = {{gReLU}},
	url = {https://www.nature.com/articles/s41592-025-02868-z},
	doi = {10.1038/s41592-025-02868-z},
	language = {en},
	number = {11},
	urldate = {2026-01-21},
	journal = {Nature Methods},
	author = {Lal, Avantika and Gunsalus, Laura and Nair, Surag and Biancalani, Tommaso and Eraslan, Gokcen},
	month = nov,
	year = {2025},
	note = {Publisher: Nature Publishing Group},
	pages = {2253--2257},
}

@article{friedman_enhancerpromoter_2024,
	title = {Enhancer–promoter specificity in gene transcription: molecular mechanisms and disease associations},
	volume = {56},
	copyright = {2024 The Author(s)},
	issn = {2092-6413},
	shorttitle = {Enhancer–promoter specificity in gene transcription},
	url = {https://www.nature.com/articles/s12276-024-01233-y},
	doi = {10.1038/s12276-024-01233-y},
	language = {en},
	number = {4},
	urldate = {2026-01-23},
	journal = {Experimental \& Molecular Medicine},
	author = {Friedman, Meyer J. and Wagner, Tobias and Lee, Haram and Rosenfeld, Michael G. and Oh, Soohwan},
	month = apr,
	year = {2024},
	note = {Publisher: Nature Publishing Group},
	pages = {772--787},
}

@article{dunbar_gene_2018,
	title = {Gene therapy comes of age},
	volume = {359},
	url = {https://www.science.org/doi/10.1126/science.aan4672},
	doi = {10.1126/science.aan4672},
	number = {6372},
	urldate = {2026-01-24},
	journal = {Science},
	author = {Dunbar, Cynthia E. and High, Katherine A. and Joung, J. Keith and Kohn, Donald B. and Ozawa, Keiya and Sadelain, Michel},
	month = jan,
	year = {2018},
	note = {Publisher: American Association for the Advancement of Science},
	pages = {eaan4672},
}

@misc{chen2025multiobjectiveguideddiscreteflowmatching,
      title={Multi-Objective-Guided Discrete Flow Matching for Controllable Biological Sequence Design}, 
      author={Tong Chen and Yinuo Zhang and Sophia Tang and Pranam Chatterjee},
      year={2025},
      eprint={2505.07086},
      archivePrefix={arXiv},
      primaryClass={cs.LG},
      url={https://arxiv.org/abs/2505.07086}, 
}

@article{
doi:10.1126/science.aad1067,
author = {Tasuku Kitada  and Breanna DiAndreth  and Brian Teague  and Ron Weiss },
title = {Programming gene and engineered-cell therapies with synthetic biology},
journal = {Science},
volume = {359},
number = {6376},
pages = {eaad1067},
year = {2018},
doi = {10.1126/science.aad1067},
URL = {https://www.science.org/doi/abs/10.1126/science.aad1067},
eprint = {https://www.science.org/doi/pdf/10.1126/science.aad1067}}

@article{kent_human_2002,
	title = {The {Human} {Genome} {Browser} at {UCSC}},
	volume = {12},
	issn = {1088-9051, 1549-5469},
	url = {http://genome.cshlp.org/content/12/6/996},
	doi = {10.1101/gr.229102},
	language = {en},
	number = {6},
	urldate = {2026-01-28},
	journal = {Genome Research},
	author = {Kent, W. James and Sugnet, Charles W. and Furey, Terrence S. and Roskin, Krishna M. and Pringle, Tom H. and Zahler, Alan M. and Haussler, {and} David},
	month = jun,
	year = {2002},
	pmid = {12045153},
	note = {Company: Cold Spring Harbor Laboratory Press
Distributor: Cold Spring Harbor Laboratory Press
Institution: Cold Spring Harbor Laboratory Press
Label: Cold Spring Harbor Laboratory Press
Publisher: Cold Spring Harbor Lab},
	pages = {996--1006},
}

@misc{stark2024dirichletflowmatchingapplications,
      title={Dirichlet Flow Matching with Applications to DNA Sequence Design}, 
      author={Hannes Stark and Bowen Jing and Chenyu Wang and Gabriele Corso and Bonnie Berger and Regina Barzilay and Tommi Jaakkola},
      year={2024},
      eprint={2402.05841},
      archivePrefix={arXiv},
      primaryClass={q-bio.BM},
      url={https://arxiv.org/abs/2402.05841}, 
}

@article{
doi:10.1126/science.adf7044,
author = {Yang Eric Li  and Sebastian Preissl  and Michael Miller  and Nicholas D. Johnson  and Zihan Wang  and Henry Jiao  and Chenxu Zhu  and Zhaoning Wang  and Yang Xie  and Olivier Poirion  and Colin Kern  and Antonio Pinto-Duarte  and Wei Tian  and Kimberly Siletti  and Nora Emerson  and Julia Osteen  and Jacinta Lucero  and Lin Lin  and Qian Yang  and Quan Zhu  and Nathan Zemke  and Sarah Espinoza  and Anna Marie Yanny  and Julie Nyhus  and Nick Dee  and Tamara Casper  and Nadiya Shapovalova  and Daniel Hirschstein  and Rebecca D. Hodge  and Sten Linnarsson  and Trygve Bakken  and Boaz Levi  and C. Dirk Keene  and Jingbo Shang  and Ed Lein  and Allen Wang  and M. Margarita Behrens  and Joseph R. Ecker  and Bing Ren },
title = {A comparative atlas of single-cell chromatin accessibility in the human brain},
journal = {Science},
volume = {382},
number = {6667},
pages = {eadf7044},
year = {2023},
doi = {10.1126/science.adf7044},
URL = {https://www.science.org/doi/abs/10.1126/science.adf7044},
eprint = {https://www.science.org/doi/pdf/10.1126/science.adf7044}}

@misc{yoon2025fastmontecarlotree,
      title={Fast Monte Carlo Tree Diffusion: 100x Speedup via Parallel Sparse Planning}, 
      author={Jaesik Yoon and Hyeonseo Cho and Yoshua Bengio and Sungjin Ahn},
      year={2025},
      eprint={2506.09498},
      archivePrefix={arXiv},
      primaryClass={cs.AI},
      url={https://arxiv.org/abs/2506.09498}, 
}

@misc{yoon2026montecarlotreediffusion,
      title={Monte Carlo Tree Diffusion for System 2 Planning}, 
      author={Jaesik Yoon and Hyeonseo Cho and Doojin Baek and Yoshua Bengio and Sungjin Ahn},
      year={2026},
      eprint={2502.07202},
      archivePrefix={arXiv},
      primaryClass={cs.AI},
      url={https://arxiv.org/abs/2502.07202}, 
}

@inproceedings{uct,
  title={Bandit based monte-carlo planning},
  author={Kocsis, Levente and Szepesv{\'a}ri, Csaba},
  booktitle={European Conference on Machine Learning},
  year={2006},
}

@article{yin_iterative_2025,
	title = {Iterative deep learning design of human enhancers exploits condensed sequence grammar to achieve cell-type specificity},
	volume = {16},
	issn = {2405-4712, 2405-4720},
	url = {https://www.cell.com/cell-systems/abstract/S2405-4712(25)00135-8},
	doi = {10.1016/j.cels.2025.101302},
	number = {7},
    year         = {2025},
	journaltitle = {Cell Systems},
	shortjournal = {cels},
	publisher = {Elsevier},
	author = {Yin, Christopher and Castillo-Hair, Sebastian and Byeon, Gun Woo and Bromley, Peter and Meuleman, Wouter and Seelig, Georg},
	urldate = {2026-05-01},
	date = {2025-07-16},
}

@article{steinegger_mmseqs2_2017,
	title = {{MMseqs2} enables sensitive protein sequence searching for the analysis of massive data sets},
	volume = {35},
	copyright = {2017 Springer Nature America, Inc.},
	issn = {1546-1696},
	url = {https://www.nature.com/articles/nbt.3988},
	doi = {10.1038/nbt.3988},
	language = {en},
	number = {11},
	urldate = {2026-05-02},
	journal = {Nature Biotechnology},
	publisher = {Nature Publishing Group},
	author = {Steinegger, Martin and Söding, Johannes},
	month = nov,
	year = {2017},
	pages = {1026--1028},
}

@misc{heusel2018ganstrainedtimescaleupdate,
      title={GANs Trained by a Two Time-Scale Update Rule Converge to a Local Nash Equilibrium}, 
      author={Martin Heusel and Hubert Ramsauer and Thomas Unterthiner and Bernhard Nessler and Sepp Hochreiter},
      year={2018},
      eprint={1706.08500},
      archivePrefix={arXiv},
      primaryClass={cs.LG},
      url={https://arxiv.org/abs/1706.08500}, 
}

@misc{gu2024mambalineartimesequencemodeling,
      title={Mamba: Linear-Time Sequence Modeling with Selective State Spaces}, 
      author={Albert Gu and Tri Dao},
      year={2024},
      eprint={2312.00752},
      archivePrefix={arXiv},
      primaryClass={cs.LG},
      url={https://arxiv.org/abs/2312.00752}, 
}

@article{dalla-torre_nucleotide_2025,
	title = {Nucleotide {Transformer}: building and evaluating robust foundation models for human genomics},
	volume = {22},
	copyright = {2024 The Author(s)},
	issn = {1548-7105},
	shorttitle = {Nucleotide {Transformer}},
	url = {https://www.nature.com/articles/s41592-024-02523-z},
	doi = {10.1038/s41592-024-02523-z},
	language = {en},
	number = {2},
	urldate = {2026-09-23},
	journal = {Nature Methods},
	publisher = {Nature Publishing Group},
	author = {Dalla-Torre, Hugo and Gonzalez, Liam and Mendoza-Revilla, Javier and Lopez Carranza, Nicolas and Grzywaczewski, Adam Henryk and Oteri, Francesco and Dallago, Christian and Trop, Evan and de Almeida, Bernardo P. and Sirelkhatim, Hassan and Richard, Guillaume and Skwark, Marcin and Beguir, Karim and Lopez, Marie and Pierrot, Thomas},
	month = feb,
	year = {2025},
	pages = {287--297},
}
